\documentclass[9pt,twocolumn,twoside]{pnas-new}

\templatetype{pnasresearcharticle} 

\title{Direct validation of dune instability theory}

\author[a,1,2]{\href{https://orcid.org/0000-0003-3142-7747}{Ping L\"u}}
\author[b,1,2]{\href{https://orcid.org/0000-0002-9556-9544}{Cl\'ement Narteau}}
\author[a]{Zhibao Dong}
\author[c]{\href{https://orcid.org/0000-0001-8975-4500}{Philippe Claudin}}
\author[b]{\href{https://orcid.org/0000-0003-1219-0641}{S\'ebastien Rodriguez}}
\author[d]{Zhishan An}
\author[b]{Laura Fernandez-Cascales}
\author[b]{\href{https://orcid.org/0000-0002-2173-5837}{Cyril Gadal}}
\author[e]{\href{https://orcid.org/0000-0002-4133-3536}{Sylvain Courrech du Pont}}

\affil[a]{School of Geography and Tourism, Shaanxi Normal University, 620 Chang'an West Avenue, Xi'an, Shaanxi 710119, China}
\affil[b]{Universit\'e de Paris, Institut de physique du Globe de Paris, CNRS, F-75005 Paris, France}
\affil[c]{Physique et M\'ecanique des Milieux H\'et\'erog\`enes, CNRS, ESPCI, PSL Research Univ, Sorbonne Univ, Universit\'e de Paris, 10 rue Vauquelin, 75005 Paris, France}
\affil[d]{Northwest Institute of Eco-Environment and Resources, Donggang West Road 320, Lanzhou, Gansu Province 730000, China}
\affil[e]{Laboratoire Mati\`ere et Syst\`eme Complexes, Universit\'e de Paris, CNRS, 10 rue Alice Domon et L\'eonie Duquet, 75205 Paris Cedex 13, France}

\leadauthor{Ping L\"u}

\significancestatement{
We report a landscape-scale experiment with controlled initial and boundary
conditions to reveal incipient dune growth under the natural action of wind.
We measured the growth rate of dunes of different wavelengths for 42~months
in a desert plot of 7,500~m$^2$. We identify an early phase in which a clear
maximum growth rate occurs for a 15~m dune spacing. A successful comparison
with dune instability theory allows to quantify the size-selection mechanism
leading to the emergence of periodic dune patterns, which can be directly
related to flow and sand transport properties. An experiment of this type
and scale is unprecedented. Its results boost confidence in the existing
theory, confirming its application to a variety of planetary landscapes from
repeated aerial/orbital imagery.
}

\authorcontributions{
P.L. and C.N. carried out all statistical data analysis, helped by C.G. for dispersion relations.
P.L., C.N., P.C. and S.C.P. designed the experimental study.
P.L. and Z.D. managed the experimental site and P.L. led the acquisition of data in the field.
Z.A. performed all laser scans.
S.R. and L.F.C. acquired and analyzed wind profiles.
C.N., P.C., S.R. and S.C.P. wrote the manuscript.
All authors participate to field work and discussed the results.}
\authordeclaration{There is no competing interest.}
\equalauthors{\textsuperscript{1}P.L. contributed equally to this work with C.N.}
\correspondingauthor{\textsuperscript{2}To whom correspondence should be addressed. E-mail: lvping@snnu.edu.cn, narteau@ipgp.fr}

\keywords{Dune instability $|$ Size-selection $|$ Sediment transport $|$ Surface winds}

\begin{abstract}
Modern dune fields are valuable sources of information for the large-scale
analysis of terrestrial and planetary environments and atmospheres, but their
study relies on understanding the small-scale dynamics that constantly
generate new dunes and reshape older ones. Here we designed a landscape-scale
experiment at the edge of the Gobi desert, China, to quantify the development
of incipient dunes under the natural action of winds. High-resolution
topographic data documenting 42~months of bedform dynamics are examined to
provide a spectral analysis of dune pattern formation. We identified two
successive phases in the process of dune growth, from the initial flat sand
bed to a meter-high periodic pattern. We focus on the initial phase, when the
linear regime of dune instability applies, and measure the growth rate of
dunes of different wavelengths. We identify the existence of a maximum growth
rate, which readily explains the mechanism by which dunes select their size,
leading to the prevalence of a 15~m-wavelength pattern. We quantitatively
compare our experimental results to the prediction of the dune instability
theory using transport and flow parameters independently measured in the
field. The remarkable agreement between theory and observations demonstrates
that the linear regime of dune growth is permanently expressed on
low-amplitude bed topography, before larger regular patterns and slip faces
eventually emerge. Our experiment underpin existing theoretical models for
the early development of eolian dunes, which can now be used to provide
reliable insights into atmospheric and surface processes on Earth and other
planetary bodies.
\end{abstract}

\dates{This manuscript was compiled on \today
\vspace{.3cm}
\newline
{\color{purple}\small  An edited version of this paper was published by PNAS un der the PNAS licence: Lü, P., Narteau, C., Dong, Z., Claudin, P., Rodriguez, S., An, Z., Fernandez-Cascales, L., Gadal, C. and du Pont, S.C., 2021. Direct validation of dune instability theory. Proceedings of the National Academy of Sciences, 118(17), \href{www.pnas.org/cgi/doi/10.1073/pnas.2024105118}{doi/10.1073/pnas.2024105118}.}
}
\doi{\url{www.pnas.org/cgi/doi/10.1073/pnas.2024105118}}

\begin{document}

\maketitle
\thispagestyle{firststyle}
\ifthenelse{\boolean{shortarticle}}{\ifthenelse{\boolean{singlecolumn}}{\abscontentformatted}{\abscontent}}{}

%
\dropcap{D}une research has always been stimulated by the question of the origin of periodic
bedforms that are ubiquitous in sand seas at all scales \cite{McKe79,Pye90,bLore}. Answering this
question has proven challenging, largely because of the variety of initial and boundary
conditions spanned by places where regular dune patterns are observed \cite{Ewin10,Gada20,Gada20a}.
In addition, considerable uncertainty exists on the efficiency of the mechanisms of dune
size-selection due to a dearth of reliable experiments on eolian dune growth. Quantitative
analysis of incipient dune formation is therefore a prerequisite for a better physical
understanding not only of the evolution of dune systems but also of transport and flow
properties responsible for the emergence of a characteristic length scale.

Underwater experiments have shown that, as soon as the flow is strong enough to transport
grains, a flat sand bed destabilizes into periodic bedforms propagating at a constant speed
\cite{Kenn63,Rich80,Cole94,Baas94,Vala05,Reff10,Cour14,Gada19}. However, wind tunnel experiments
and Ralph Bagnold's attempts in the field to create artificial dunes have failed because the
initial sand  piles were not large enough \cite{bBagn,Andr02a}. There is indeed a minimum
length-scale for the formation of dunes, which has been estimated to be of the order of
$10\;{\rm m}$ in eolian systems on Earth based on the smallest wavelength of the superimposed
bedforms observed on the flanks of large dunes \cite{Andr05,Part07,Nart09}. After 20 years of
intensive research, this characteristic length-scale is assumed to be regulated by the balance
between a destabilizing process associated with the turbulent flow response to the topography
and a stabilizing process due to transport inertia \cite{Char13}.

Fluid mechanics studies have shown that the flow above low dunes can be decomposed into an inner
and an outer layer \cite{Jack75}. In the outer layer, the flow perturbation is in phase with
topography (i.e., wind speed is maximum above the dune crest). In the inner layer (the thin
region adjacent to the bed), the effect of friction causes the wind velocity to be phase-advanced
\cite{Hunt88,Clau13}. The theory predicts that this layer has a thickness, $l$, that shows a
positive dependence on the wavelength, $\lambda$, of the bedforms. Within this layer, the maximum
(minimum) wind speed is upwind of the maximum (minimum) of topography. For a sinusoidal topography,
this perturbation takes the form of a phase shift, $\varphi_{\rm b}$, which is usually described
with two terms, $\cos(kx+\varphi_{\rm b})=A\cos(kx)+B\cos(kx+\pi/2)$, where $k=2\pi/\lambda$ is the
dune wave number. The in-phase term has a weight coefficient $A$. The other, in
phase quadrature, has a weight coefficient $B$. Then, the upwind shift between the wind speed and
the bed topography can be expressed as $\arctan(B/A)/k$ . Since the dune crest is a zone of
decreasing wind speed, and therefore of sand deposition, this spatial shift is the essential
ingredient for the initial growth of dunes \cite{Kroy02}. Nevertheless, the length scale associated
with this spatial shift does not alone promote the dominance of any specific wavelength. Wavelength
selection requires another antagonistic process, which relies on sand transport properties.

As sand flux adjusts to the aforementioned spatial change in wind speed, there is a spatial lag
between the loci of maximum wind speed and maximum transport rate \cite{Saue01,Paht13,Andr10}.
This length scale is known as the saturation length $l_{\rm sat}$, the distance required for the
sand flux to reach saturation. The combined effect of this downwind shift in sand transport and
the upwind shift in wind speed eventually acts as a size-selection mechanism. The minimum length
scale for the formation of dunes arises at a characteristic wave
number, $k_0$, defined as the wave number for which these downwind and upwind shifts cancel each
other out. At longer wavelengths ($k<k_0$), the crest is a zone of net deposition such that the
dune grows. However, larger bedforms have longer response times which means that the growth rate
also tends to zero in the limit $k\to0$. There is therefore a characteristic wave number,
$k_{\rm max}$, for which the growth rate is maximum. It determines the most unstable wavelength,
$\lambda_{\rm max}$, and the characteristic length-scale for the emergence of periodic dunes.

Linear stability analysis of flat sand beds sheared by a fluid flow provide the analytical
expression for the growth rates $\sigma(k)$ of sinusoidal bed perturbations over the whole
range of possible wavelengths
\cite{Kenn63,Rich80,Andr02a,Andr12,Clau06,Colo04,Deva10,Lagr03,Dura19,Four10,Gada19,Gada20a}.
These stability analyses are by definition restricted to the linear regime of incipient dune
growth, the period during which the amplitude of each mode (wavelength) grows exponentially
and independently from one another. In this linear regime, a dispersion relation gives the
growth rate of each mode as a function of the wave number. According to the flow and transport
processes described above, theoretical dispersion relation can be expressed as
\begin{equation}
\sigma(k)=Q\,k^2\,\frac{B-A\,k\,l_{\rm sat}}{1+\left( k\,l_{\rm sat} \right)^2},
\label{eq: disp1}
\end{equation}
where $Q$ is the mean sand flux, which can be derived from wind data and the threshold shear
velocity, $u_{\rm th}$, for the transport of sand grains \cite{Char13}. The numerator not only
governs the initial dune growth (i.e., $\sigma(k)>0$ for $k<k_0$ and $\sigma(k)<0$ for $k>k_0$,
where $k_0=B/(A\,l_{\rm sat})$) but also provides the essential
mechanism for dune-size selection during the linear regime. Later, as dune aspect ratios
become larger, dune dynamics enter a non-linear growth regime dominated by collisions and
interactions between dunes of different sizes \cite{Ewin06,Part06,Vala11,Gao15b,Day18},
and this dispersion relation no longer applies.

Focusing only on well-established periodic dune patterns, the existence of a characteristic
wavelength for dune formation has been validated in various eolian environments using tools
of comparative planetology \cite{Clau06} and in different places on Earth using superimposed
bedforms \cite{Andr05} as well as incipient dunes in ephemeral channels \cite{Delo20} or
at the border of dune fields \cite{Gada20}.  However, the dependence of the growth rate on
dune size and the underlying size-selection mechanism associated with the emergence of
periodic dunes have never been observed and quantified in a natural eolian environment,
nor in wind tunnels. To bridge this gap, we performed a long-term field experiment to measure,
in the same desertic area, all of the parameters involved in the physics of dune growth. We
present here a comprehensive description of incipient dune growth under the natural action of
wind, which is directly confronted to the dune instability theory.

\section*{A landscape-scale experiment}
\begin{figure*}
\centering
\includegraphics[width=17.8cm]{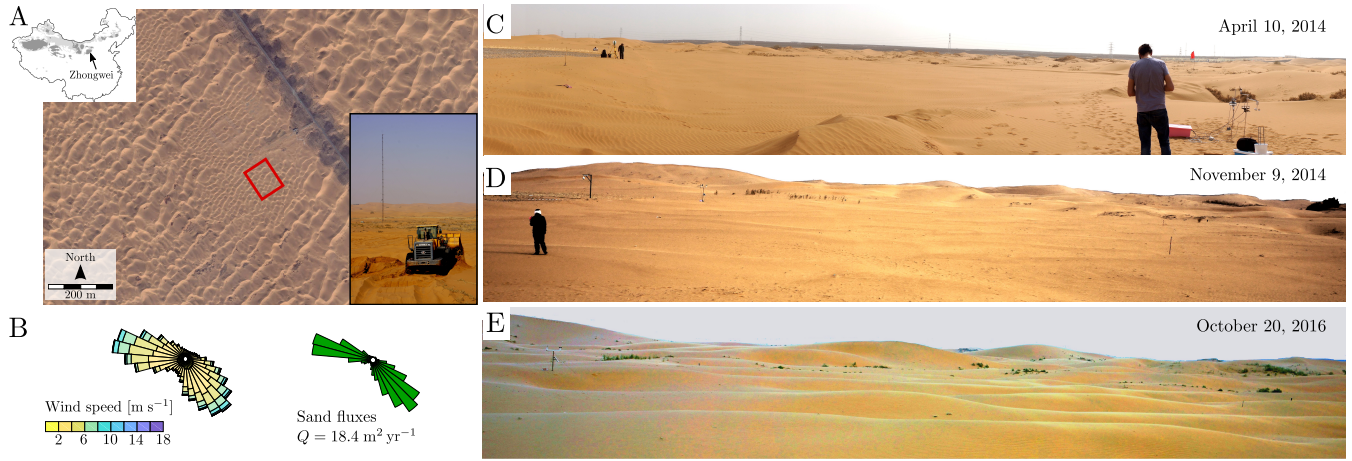}
\caption{
A landscape-scale experiment on incipient dune growth.
({\em A})
The experimental site in the Tengger desert
(37$^{\rm o}$33'38.14''N,
105$^{\rm o}$2'0.76''E).
The red square shows location of the flat sand bed experiment.
({\em B})
Wind and sand flux roses from January 2014 to November 2017.
({\em C})
View of the flat sand bed experiment in April 2014.
({\em D})
Incipient dunes at the end of the linear phase of dune growth in November 2014.
({\em E})
Mature dunes during the non-linear phase of dune growth in October 2016.
}
\label{fig: fig1}
\end{figure*}

In our landscape-scale experiments, we investigated dune dynamics at a scale of hundred
of meters using controlled initial conditions and continuous monitoring of environmental
conditions \cite{Ping14}. These experiments started in 2009 in the Tengger Desert at the
southeastern edge of the Gobi basin in China (Fig.~\ref{fig: fig1}A). This area is exposed
to a bidirectional wind regime (Fig.~\ref{fig: fig1}B). The primary wind blows from the
northwest mainly in the spring when the Siberian high-pressure system weakens. In summer,
the easterly wind of the east-Asian monsoon dominates. Since November 2013, wind data
from a local airport located $10\;{\rm km}$ east of the field site are synchronized with
the wind data from the local meteorological tower and a $2\;{\rm m}$ high wind tower
located on the side of our new experimental plot (Fig.~S1).

The experiments dedicated to incipient dune growth have been conducted from April 2014 to
November 2017. Pre-existing dunes were leveled on April 9, 2014, to form a flat rectangular
bed $100\,{\rm m}$ long and $75\,{\rm m}$ wide (Fig.~\ref{fig: fig1}C).  The long axis of this
rectangular area is aligned with the direction of the primary wind. We monitored dune growth
over the following $42$ months (Fig.~\ref{fig: fig1}~D and E) through a series of 20 topographic
surveys using a ground-based laser scanner. To get a better resolution on the early stage of
dune growth and to account for windy periods, these topographic data are not regularly
distributed in time and are more frequent in 2014 as well as in the spring and fall of each
year. To compare datasets from different scans, we installed a reference system of concrete
posts over the entire experimental dune field (Fig.~S9). During each topographic survey, we
scanned the experimental plot from four different locations. Depending on the local slopes, the
density of points varied from $472$ to $2368$~points/m$^2$ with a centimetric height accuracy
throughout the duration of the experiment.

\section*{Results}
\subsection*{Sand transport properties}
\begin{figure*}
\centering
\includegraphics[width=8.7cm]{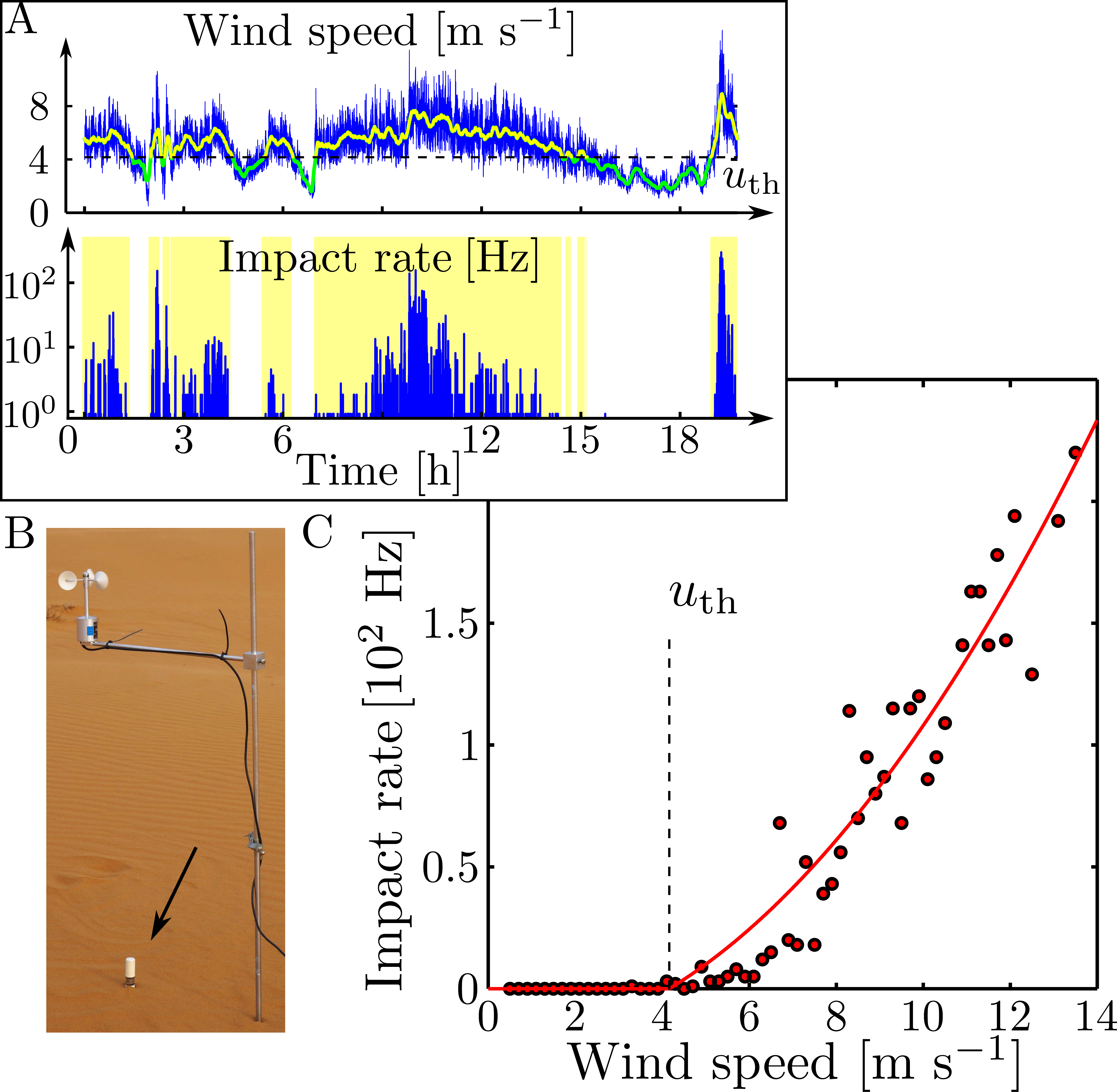}
\caption{
The threshold wind speed for sand transport.
({\em A})
Grain impact rate and wind speed with respect to time.
({\em B})
Picture of the impact sensor placed above a flat sand and below a cup anemometer.
({\em C})
Relationship between grain impact rate and wind speed $u$ measured at a height of $1.3\,{\rm m}$.
The solid line is the best fit of $(u^2-u_{\rm th}^2)$ to the data with a critical entrainment threshold
$u_{\rm th}=4.16\;{\rm m\,s}^{-1}$. Periods above the transport threshold are highlighted in yellow in
Fig.~2A.
}
\label{fig: fig2}
\end{figure*}

To measure the threshold wind speed for sand movement and estimate the local sand flux from
wind data, we characterized sand transport in the field in a dedicated set of short-term
experiments. As in most of the Tengger Desert, the mean grain size in the experimental area is
of about $190\;\mu{\rm m}$ (Fig.~S2). Using an impact sensor placed above a flat sand bed, we
monitored saltation activity under winds of varying strength \cite{Stou97}. As shown in
Fig.~\ref{fig: fig2}, the empirical relation between wind speed and the impact rate yields an
estimated threshold wind speed for aerodynamic entrainment of sand grains,
$u_{\rm th}=0.23\pm0.04\;{\rm m\,s}^{-1}$. Combining this threshold value with local wind data,
we calculate the saturated sand flux on a flat sand bed using the eolian transport law of
Ungar and Haff \cite{Unga87,Dura11,Cour14}. From January 2014 to November 2017, the mean flux
is $Q=18.4\pm4.2\;{\rm m^2\,yr}^{-1}$ (Tab.~S1).

\begin{figure*}
\centering
\includegraphics[width=17.8cm]{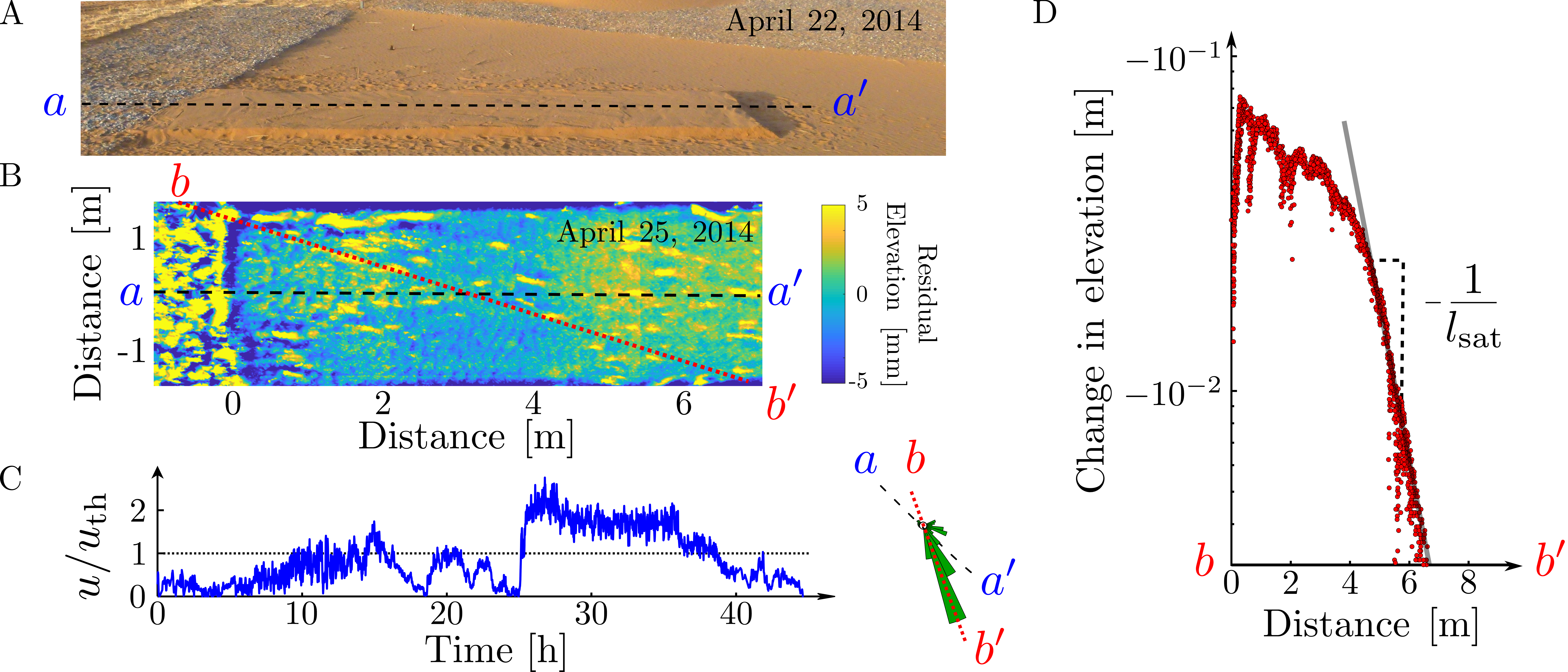}
\caption{The saturation length experiment.
({\em A})
The initial sand slab prepared and scanned on April 22, 2014. Placed downstream of a non-erodible bed
composed of gravels, it has a length of $12\;{\rm m}$, a width of $3\;{\rm m}$ and a main axis $aa'$
aligned in the northwest-southeast direction, parallel to the orientation of the prevailing wind.
({\em B})
The residual elevation of the sand slab on April 25, 2014 after the passage of a storm event. It is
the difference between the raw elevation map and a smoothed elevation map using a radius of $20\;{\rm cm}$.
({\em C})
Wind speed and sand flux rose between the two scans.
({\em D})
Difference in surface elevation along the profile $bb'$ between the two scans. Aligned with the resultant
transport direction, this profile is selected from the orientation of ripples in $B$ and the sand flux rose
in $C$. The negative values show the net erosion along the entire profile, $\Delta H \sim - \exp(x/l_{\rm sat})$.
An exponential fit to the data gives $l_{\rm sat}=0.95\;{\rm m}$ (Supplementary Information).
}
\label{fig: fig3}
\end{figure*}

To estimate the saturation length, $l_{\rm sat}$, we constructed a flat bed armored with coarse
gravels and cobbles, which acted as a sand trap. At a time without active transport, we build
a $12\,{\rm m}$ long, $3\,{\rm m}$ wide and $20\,{\rm cm}$ high flat sand berm immediatly
downwind of the coarse bed, as determined from the direction of the prevailing wind
(Fig.~\ref{fig: fig3}A). We measure the surface elevation of this sand berm using the terrestrial
laser scanner before and after a wind event (Fig.~\ref{fig: fig3}~B and C). The difference in
topography along the  wind direction gives the mean transport rate profile for this time interval
(Figs.~S3 and S4). Fig.~\ref{fig: fig3}D shows a net erosion on the whole sand slab, with an
amplitude that dampens with respect to the downwind distance from the non-erodible bed. These
observations indicate an increasing sand flux converging toward its saturated value. Assuming an
exponential relaxation of the sand flux \cite{Andr10}, we obtain $l_{\rm sat}=0.95\pm 0.2 \,{\rm m}$.

\subsection*{Flow perturbation over low sinusoidal bedforms}
\begin{figure*}
\centering
\includegraphics[width=17.8cm]{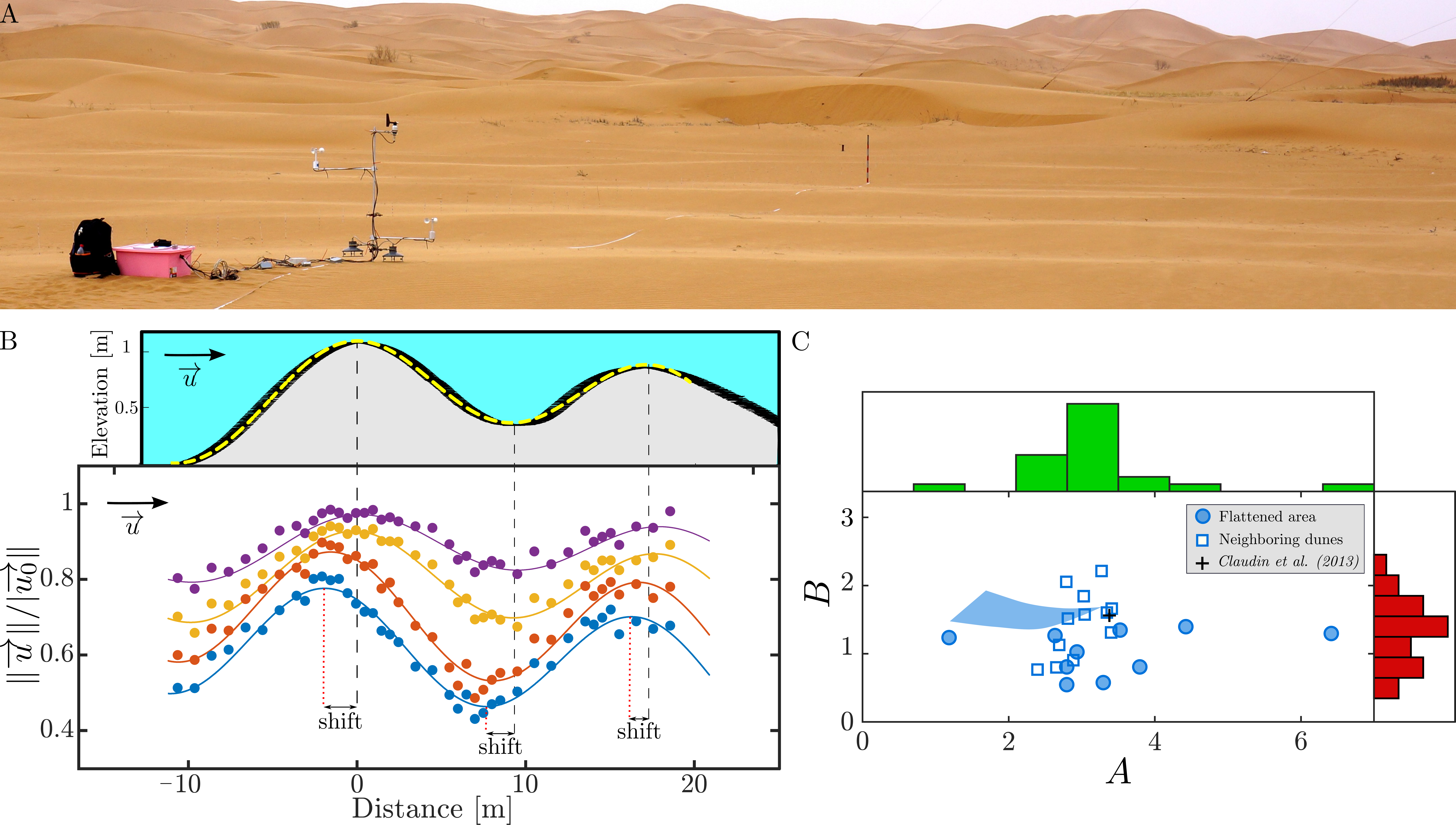}
\caption{
The upwind velocity shift on low sinusoidal bedforms.
({\em A})
The mobile anemometer mast, with anemometers located at heights of $4$ and $12\;{\rm cm}$ in
the inner layer and at heights of $50$ and $100\;{\rm cm}$ in the outer layer. The decameter
aligned with the mean wind direction gives the direction followed during the successive
measurements on sinusoidal incipient dunes.
({\em B})
Elevation and normalized wind speed according to distance along the dune profile (blue, red,
yellow and purple for anemometers at $4$, $12$, $50$ and $100\;{\rm cm}$, respectively). The
normalized wind speed is the wind speed measured along the profile divided by the one recorded
by the reference anemometer. Note the upwind shift in wind speed in the inner layer near the
bed, but not in the outer layer above. Dashed lines lines show the two dune crests and the
trough along the elevation profile. Dotted lines show the maximum and minimum wind speed in
the inner layer.
({\em C})
Values and distribution of aerodynamic parameters $A$ and $B$ measured during dune growth
(circles) and on neighboring dunes (squares). The blue area shows the best fit values of
$A$ and $B$ to the experimental dispersion relation shown in Fig.~\ref{fig: fig6}B for
$0.6<l_{\rm sat}<1.2$.
}
\label{fig: fig4}
\end{figure*}

To estimate the values of the aerodynamic parameters $A$ and $B$ used in Eq.~\ref{eq: disp1},
we measured the flow properties in October~2013 on dunes of small amplitude in the neighborhood
of the experimental site. After flattening, the same measurements for the estimation of the
phase shift between wind speed and topography were repeated in November 2014 as well as
in April and November 2015 over the incipient bedforms of our experimental plot. We recorded
wind speed at heights of $4$, $12$, $50$, and $100\,{\rm cm}$ above the bed by moving an anemometer
mast upwind of a known elevation dune profile, which include in most cases a succession
of crests and troughs (Figs.~\ref{fig: fig4}A and S5). The density of the measurements is higher
near the crests and troughs in order to determine more precisely the position of the maximum and
minimum wind speeds. Each measurement lasted at least 10~min with a sampling rate of $1$~Hz. The wind
speeds measured at different heights were synchronized and normalized by the wind speed measured
at one meter high by a reference cup anemometer located at the top of a larger dune in the
vicinity, which allowed us to reconstruct the perturbation in wind speed on low dunes at different
stages of dune growth.

As predicted in the limit of low sinusoidal bedforms, Fig.~\ref{fig: fig4}B shows that the wind
perturbation at all heights reflect the topography of the underlying incipient dunes, both in
amplitude and wavelength. The amplitude of perturbation in wind speed decreases with height above
the bed. More importantly,
there is always an upwind shift in wind speed for the two bottom anemometers at heights of $4$ and
$12\,{\rm cm}$, but not for the two top anemometers at heights of $50$ and $100\,{\rm cm}$. This
indicates that, on incipient dunes in our experimental area, the thickness $l$ of the inner-layer
is between $12$ and $50\;{\rm cm}$. Within this inner layer, using the upwind shift
($\approx 1 \,{\rm m}$) and the amplitude of the perturbation in wind speed recorded by the two
bottom anemometers, we get $A=3 \pm 1$ and $B=1.5 \pm 0.5$ for dune aspect ratios varying from
$0.012$ to $0.025$ (Fig.~\ref{fig: fig4}C and Tab.~S2).

\subsection*{Emergence of a periodic dune pattern}
\begin{figure*}
\centering
\includegraphics[width=16.0cm]{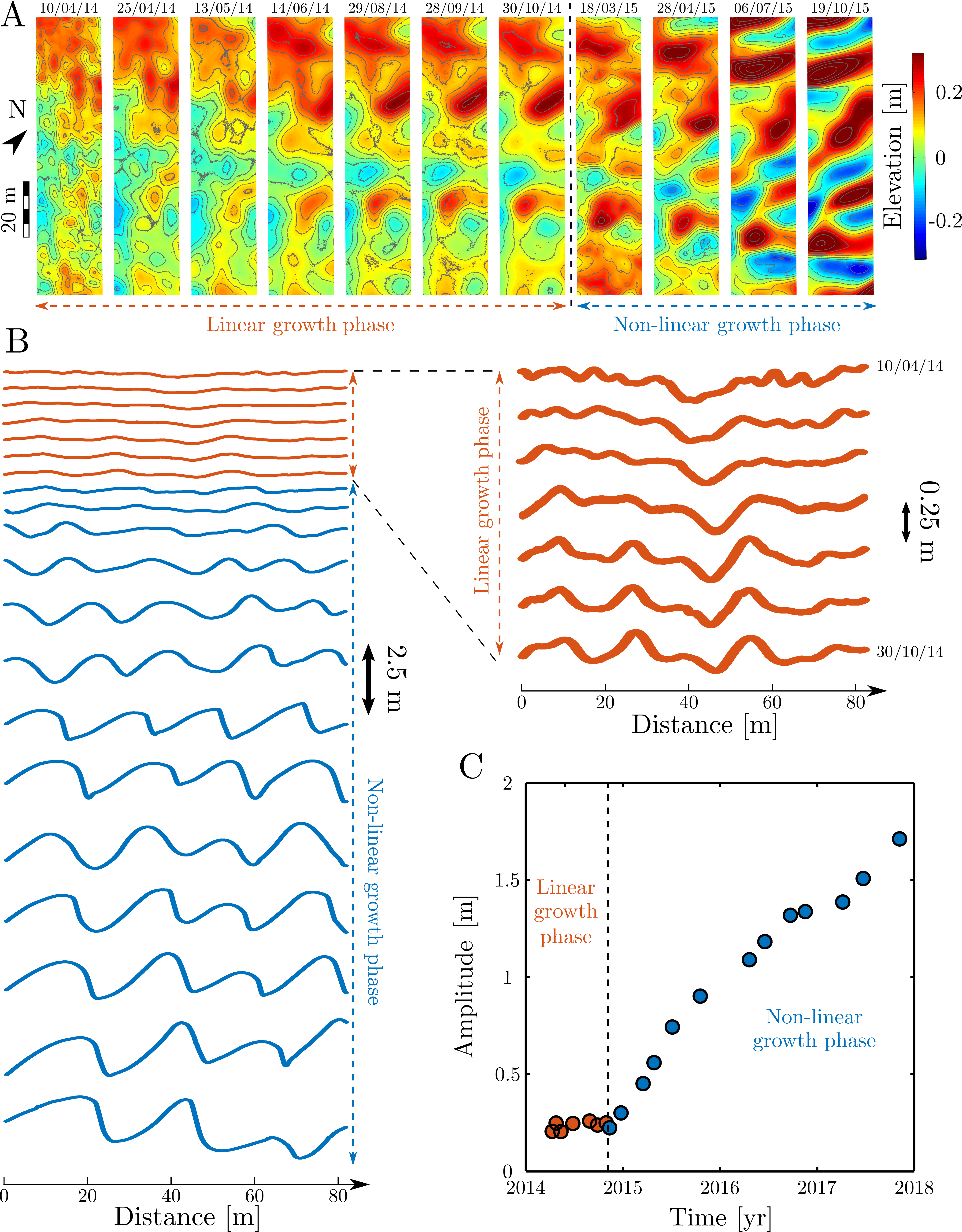}
\caption{
Dune growth in the flat sand bed experiment.
({\em A})
Elevation maps during incipient dune growth from April 10, 2014 to October 19, 2015.
({\em B})
Detrended elevation profiles along the same transect from April 10, 2014 to November 7, 2017
obtained by averaging elevation data points over a $0.4$~m wide band. Zooms on the elevation
profiles from April 10, 2014 to October 30, 2014 show the evolution of topography during the
linear phase of dune growth.
({\em C})
The mean amplitude $2\sqrt{2}(\langle h^2\rangle-\langle h\rangle^2)^{1/2}$ of bedforms with
respect to time. Colors and dashed lines are used to separate the linear (orange) and the
non-linear (blue) phases of dune growth.
}
\label{fig: fig5}
\end{figure*}

Within the experimental plot, to avoid disturbances from the surrounding bedforms, we select a
central rectangular area with a width of 48~m and a length of 82~m (red square in
Fig.~\ref{fig: fig1}A). The long side of this rectangle is oriented Northwest-Southeast to align
with the prevailing transport direction. We remove the mean slope of this rectangular area by
fitting a plane to the elevation data. Throughout the experiment, this plane maintained a
gentle southwest-facing slope as observed after the flattening of the dune field. The residual
topography is shown for different times in Fig.~\ref{fig: fig5}A. Within the observation area,
we chose to follow the time evolution of elevation along $34$ parallel transects with a constant
spacing of $1.4\;{\rm m}$. These transects are oriented perpendicularly to the final dune
orientation observed in November 2017. Fig.~\ref{fig: fig5}B shows the elevation profiles with
respect to time for a given transect. Over the 42 months of the experiment, the amplitude of the
dunes increases by two orders of magnitude from a few centimeters to a few meters. Whereas no
periodic pattern is discernible after flattening, a characteristic wavelength of $15$~m emerges
over the first few months of the experiment (Figs.~\ref{fig: fig1}D and \ref{fig: fig1}E).

The variation of the mean amplitude of the dune pattern, defined as the root‐mean‐square of the
topography, is not homogeneous over time and displays a sudden change in rate at the end of 2014
(Fig.~\ref{fig: fig5}C). Before, from April to October 2014, this amplitude stays almost constant.
During this time period, despite a smoother topography, there is nothing to suggest that dunes
will appear at a specific wavelength. Starting in November 2014, the surface elevation
exhibits a periodic dune pattern with marked crestlines and a northeast-southwest orientation.
Then, the mean amplitude increases significantly at a constant rate of $0.5\;{\rm m}\;{\rm yr}^{-1}$
for three years (Fig.~\ref{fig: fig5}C). We found that the transition between these two different
stages of dune growth occurs for a mean slope of the order of 0.03 (Fig.~S6), when dune crests and
slip faces emerge and begin to spatially organize throughout the experimental plot (Fig.~S7).
Steeper slopes highlights the increase in dune aspect ratio \cite{Phil19,Gada20a}, which is the
main control parameter for areodynamic non-linearities. Hence, we ascribe the two different stages
to the linear and the non-linear phases of the dune growth instability.

\subsection*{Experimental dispersion relation for incipient dune growth}
\begin{figure*}
\centering
\includegraphics[width=11.4cm]{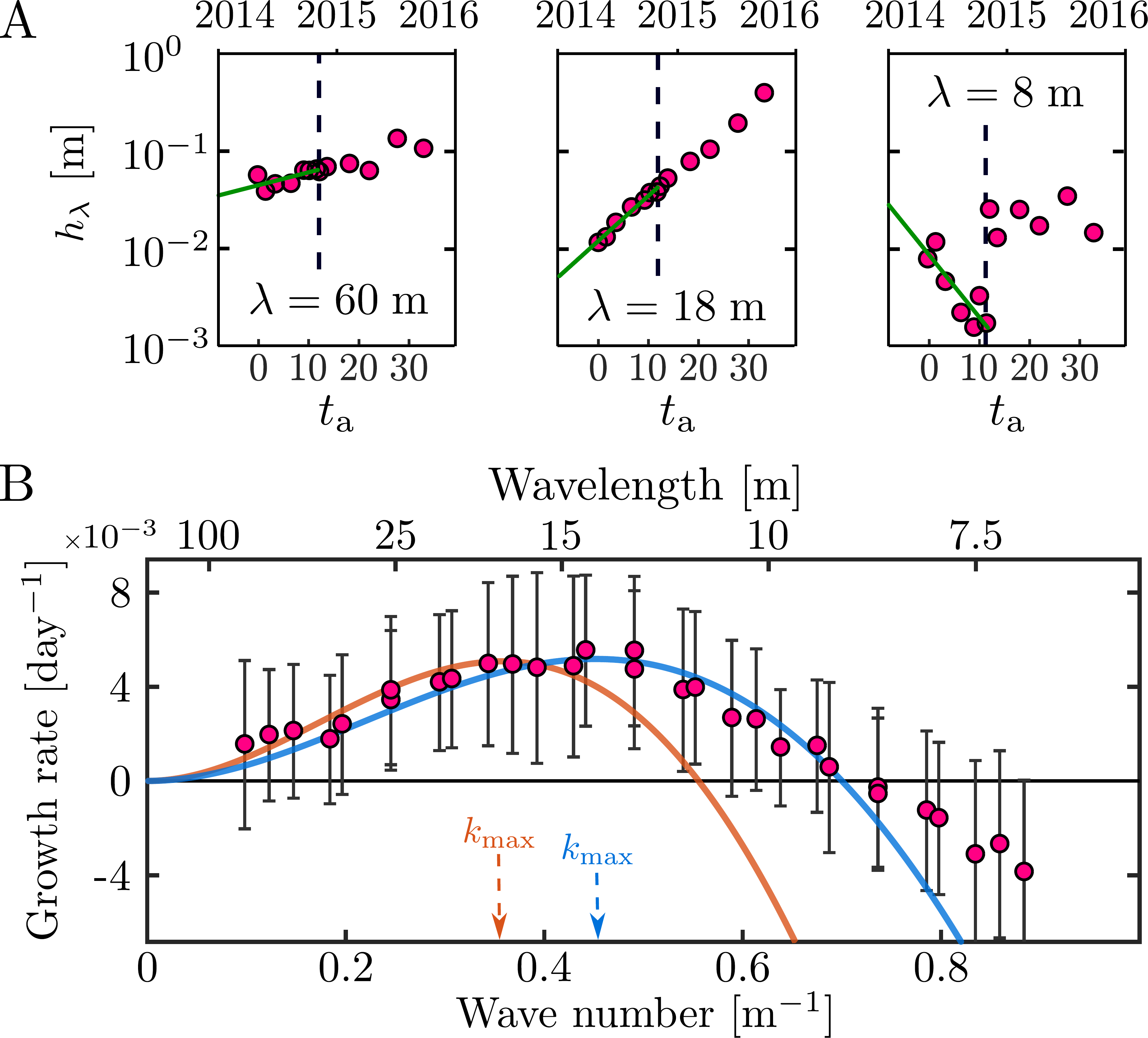}
\caption{
The dispersion relation of the dune instability in the flat sand bed experiment.
({\em A})
Amplitude of $60$, $18$ and $8\,{\rm m}$ wavelengths with respect to the dimensionless transport
time scale $t_{\rm a}$; $t_{\rm a}$ is set to $0$ on April 10, 2014. Two different regimes are
observed before and after $t_a=10.7$, October 30, 2014 (dashed lines). There are associated with
the linear and non-linear phases of dune growth. The growth rate of each wavelength (green lines)
is determined by an exponential fit performed during the linear phase ($0\leq t_a \leq 10.7$).
({\em B})
The experimental dispersion relation for the linear regime of the dune instability, i.e., the
growth rate with respect to the wave number during the linear phase (dots). Errorbars show mean
and standard deviation derived from the 34 independent transects. Solid lines are dispersion
relations using Eq.~\ref{eq: disp1} and the best fit to the data
$\{l_{\rm sat},\, A,\, B\}=\{0.7,\,1.96,\,0.96\}$ (blue),
or the same parameters measured independently in the field
$\{l_{\rm sat},\, A,\, B\}=\{0.95,\,3,\,1.5\}$ (orange).
The most unstable wavelengths $\lambda_{\rm max}=2\pi/k_{\rm max}$ are equal to $14.6$ and
$18.5\;{\rm m}$, respectively.
}
\label{fig: fig6}
\end{figure*}

Elevation profiles were repeatedly acquired at 20 different times at each of our 34 transect
locations. We performed a spectral decomposition of these elevation profiles to isolate the
contribution of individual modes (i.e., wavelengths) to the overall topography. Fig.~\ref{fig: fig6}A
shows the amplitude of three different modes as a function of a dimensionless transport time
scale, which is set to zero after flattening and is then incremented over time proportionally
to sand flux (Fig.~S8). Once again, abrupt changes in growth rate are observed in November 2014.
Before, the amplitude of the different modes is found to vary exponentially with time, which is
consistent with the linear regime of the dune instability.

For length scales ranging from $5$ to $60\,{\rm m}$, we measure the exponential growth rate of
each mode from April to October 2014 using the dimensionless time (Figs.~S9-S11). It is positive for
large wavelengths, negative for small wavelengths, and reaches a maximum value for an intermediate
wavelength of approximately $15\,{\rm m}$. Plotting the growth rate of the different modes as a
function of their wave number $k$, we obtain the experimental dispersion relation of the dune
instability (Fig.~\ref{fig: fig6}B). The consistent variation of these growth rates over the entire
range of length scales, the presence of clear maximum and the continuous trend from unstable (i.e.,
growing  waves, $\sigma>0$) to stable regimes (i.e., decaying waves, $\sigma<0$) reflect most of
the behaviors usually predicted during the linear phase of dune growth. Most importantly, it
provides the first eolian experimental evidence of the difference in growth rates of nascent dunes
of various wavelengths when they are not large enough to generate flow recirculations.

Finally, we test the consistency of the theoretical prediction of the linear stability analysis
by comparing the dispersion relations given by Eq.~\ref{eq: disp1} to the experimental dispersion
relation derived only from topographic data in the field. We either use fitted values for the
parameters $A$, $B$, and $l_{\rm sat}$, or we set them to the values we have independently measured
in the field (Figs.~\ref{fig: fig2}-\ref{fig: fig4}). In both cases, we find a good quantitative
agreement between the experimental and the theoretical dispersion relations (Figs.~\ref{fig: fig6}B
and S12-S13). As shown by the shaded area in Fig.~\ref{fig: fig4}C and a $B/A$ ratio close to
$0.5$ in both cases, the difference lies mainly in the sensitivity of the dispersion relation to
the $l_{\rm sat}$ value.

\section*{Discussion}
In order to meet the challenge of comparing and evaluating theoretical models of dune growth with
observational data, the landscape-scale experiment conducted in an active dune field under the
natural action of wind yield a unique set of new quantitative relationships. By removing
uncertainties about boundary and initial conditions, we verify that dunes can emerge from a flat
sand bed \cite{Coop58,Kocu92} and validate the theory behind this dune growth mechanism. We
elucidate the origin of periodic bedforms, showing the wavelength selection as dunes increase in
height. Meanwhile, we highlight the inherent benefits of combining field observations with theory
to derive information about the sediment transport and flow properties from the morphodynamics of
incipient dunes.

Through experimental evidence in the field, this direct validation of dune instability theory
transforms a working hypothesis into a comprehensive and operational model for the origin and
initial growth of dune patterns. From a dispersion relation based on measurable physical quantities,
such a model reveals the consistency of the size-selection mechanisms leading to the development of
spatial structures in dune fields. As soon as a granular bed sheared by a flow leads to such an
instability and the emergence of periodic dunes, their regular spacing corresponds to the most unstable mode,
determining the balance between the downwind shift in sand transport (i.e., the saturation length)
and the upwind shift in wind speed . The neutral mode ($\sigma(k_0)=0$) sets the minimum length
scale for the formation of dunes ($2\pi/k_0\approx 9\,{\rm m}$ in this area of the Tengger Desert),
and is directly estimated from observations and the experimental dispersion relation.

We provide an original dataset for an in-depth understanding of the linear phase of the
dune instability, when the growth rates of the different modes evolve independently from each other.
We find that this linear phase is at work from the earliest stage of dune growth, as soon as sand
transport starts even when, at first glance, no regular structure seems to be in place. It takes time
for the most unstable wavelength to prevail over all the other modes that contribute to the sand bed
topography. When periodic dunes are observed, the non-linear phase has already taken over, aerodynamic
non-linearities have developed and the different modes interact with each other to lead to pattern
coarsening \cite{Vala11,Gao15b}. We show here that the continuous transition from the linear to the
non-linear regimes is controlled by the dune aspect-ratio for remarkably low values ($\approx 0.03$),
consistent with laboratory measurements on sinusoidal beds \cite{Char13}. This transition coincides
with the spatial organisation of bedforms according to the alignment of mature dunes. For mean slope
values of $0.07$ ($4^{\rm o}$, typical of the flattest dune slopes), the transition is completed and
it is no longer possible to differentiate between the growth rates of the different modes.
Nevertheless, the most unstable mode can still be observed to provide relevant length and time scales
of the dune instability.

Another objective of our series of measurements was to estimate changes in the upwind velocity
shift at different stages of dune growth. A decreasing value of this phase shift would reduce the
deposition rate at the crest and could naturally be associated with the selection of the steady
aspect-ratio of dunes. This phenomenon can be expressed through a decrease in the values of the parameters $A$
and $B$ during the transition from the linear to the non-linear regime and the development of
aerodynamic non-linearities. Thus, in addition to their weak (logarithmic) positive dependences on
$kz_{\rm s}$, where $z_{\rm s}$ is the aerodynamic roughness, a weakly non-linear expansion of
the flow predicts that the parameters $A$ and $B$ have also a quadratic negative dependence on dune
aspect-ratio \cite{Andr09}. However, with our measurement precision and the range of measured dune
aspect-ratio, we could not observe any systematic dependence of the value of the parameters $A$ and
$B$ on dune growth, which remains a challenge that goes beyond the mechanism of size selection.

Landscape-scale experiments at the scale of thousands of ${\rm m}^2$ and over many years are still
extremely rare in the field of geomorphology and surface processes. Thanks to a series of
high-resolution elevation maps, this study emphasizes that we can now achieve the precision needed
to analyze the linear phase of eolian dune growth, a short but fundamental period during which
regular dune patterns may still be imperceptible. More generally, our experiments bridge the gap
between theoretical physics and geophysical surveys and can translate into concrete new research
avenues across scientific domains. Instability theory applies to many natural systems, especially
to examine the relationships between flow and surface properties on solid bedforms, icy or rocky,
controlled by chemical reactions or phase changes \cite{Cohe20,Bord20}. For all these natural
systems, we show here that, when the technological step in data acquisition is taken
\cite{Badd18,Davi20}, field studies can be carried out to test theoretical outputs against
observations and to develop better forecasts of landscape morphogenesis.

The dispersion relation gives the range of wavelengths over which nascent dunes are likely to be
observed. Then, in addition to the most unstable mode, we can now evaluate how other modes with
lower growth rates contribute to the distribution of dune sizes encountered in dune fields. As it
was done before using only the emerging dune pattern \cite{Andr05,Ping14,Gada20a,Delo20}, the
evolution of bedforms over a wide range of length scales gives a more complete framework to derive
information about the nature of the sediment and the atmospheric wind regimes from the morphodynamics
of incipient dunes. This includes air and sediment densities, grain size, and the mean wind strength,
which together determine the threshold wind speed for eolian transport.

Based on our comprehensive assessment of the dune instability, we showed here that successive
topographic surveys and spectral analysis can be combined to characterize different environments on
Earth and other planetary bodies where dunes actively participate in landscape dynamics
\cite{Clau06,Luca14,Lapo16,Fern18,Telf18,Dura19,Davi20,Lore20}. We provide quantitative evidence that
such an inverse problem can be solved over relatively short time scales relative to those involved in
the dynamics of major dune systems. This approach can be used to remotely assess sediment and
atmospheric quantities. It is particularly relevant on planetary and cometary surfaces where bedforms
can grow under exotic conditions with unusual parameter values \cite{bLore,Jia17}. The dune
instability also occur on larger structures, and this is especially visible when the wind suddenly
changes direction. An accurate, high-frequency topographic analysis based on the instability theory
could then be performed at a local scale to document wind speed and direction as well as the duration
of wind events.

\matmethods{
The Supplementary Information includes wind data (Fig.~S1), the grain size distribution (Fig.~S2)
as well as the equations and parameter values used to compute sand transport (Tab.~S1). It also
provides a detailed description of the field experiments and the equations used to estimate the
saturation length, $l_{\rm sat}$ (Figs.~S3 and S4) and the aerodynamic parameters $A$ and $B$
(Fig.~S5 and Tab.~S2). Then, it describes the solution strategy employed to obtain the dispersion
relation from the topographic data (Figs.~S6-S13).
}

\showmatmethods{} 

\acknow{We acknowledge financial support from
the National Natural Science Foundation of China (41871011 and 41930641) ,
the UnivEarthS LabEx program (ANR-10-LABX-0023),
the IdEx Universit\'e de Paris (ANR-18-IDEX-0001),
the French National Research Agency (ANR-12-BS05-001-03/EXO-DUNES, ANR-17-CE01-0014/SONO)
the National Science Center of Poland (grant 2016/23/B/ST10/01700),
and the French Chinese International laboratory SALADYN.}

\showacknow{} 


\newpage

%

\newpage

%

\newpage

%

\newpage

%

\newpage

%

\newpage

%

\end{document}


\sidecaptionvpos{figure}{m}
\sidecaptionvpos{table}{m}






\begin{center}
{\bf {\LARGE
Direct validation of dune instability theory
\vskip 0.9cm}}
{\Large --- Supplementary Information ---
}
\end{center}
\vskip 1.5cm
{\large
\centerline{Ping L\"u$^{1,*}$, Cl\'ement Narteau$^{2,*}$, Zhibao Dong$^{1}$,}
\centerline{Philippe Claudin$^{3}$, S\'ebastien Rodriguez$^{2}$, Zhishan An$^{4}$,}
\centerline{Laura Fernandez-Cascales$^{2}$, Cyril Gadal$^{2}$, Sylvain Courrech du Pont$^{5}$}
}
\vskip 1.5cm
\begin{center}
\begin{minipage}{0.72\linewidth}
{\small
\begin{itemize}
\item[$^{1}$]
School of Geography and Tourism, Shaanxi Normal University, 620 Chang'an West Avenue, Xi'an,
Shaanxi 710119, China.
\item[$^{2}$]
Universit\'e de Paris, Institut de physique du globe de Paris, CNRS, F-75005 Paris, France.
\item[$^{3}$]
Physique et M\'ecanique des Milieux H\'et\'erog\`enes,
UMR 7636 CNRS, ESPCI PSL Research Univ, Sorbonne Univ, Universit\'e de Paris,
10 rue Vauquelin, 75005 Paris, France.
\item[$^{4}$]
Northwest Institute of Eco-Environment and Resources
Donggang West Road 320, Lanzhou, Gansu Province 730000, China.
\item[$^{5}$]
Laboratoire Mati\`ere et Syst\`eme Complexes,
Universit\'e de Paris.
UMR 7057 CNRS, B\^atiment Condorcet,
10 rue Alice Domon et L\'eonie Duquet, 75205 Paris Cedex 13, France.
\end{itemize}
}
\end{minipage}
\end{center}

\vskip 8.0cm
\rule{0.38\textwidth}{.4pt}
\vskip 0.0cm
$^*$To whom correspondence should be addressed. E-mail: {\tt lvping@snnu.edu.cn}, {\tt narteau@ipgp.fr}

\newpage

\vskip 2.0cm
\tableofcontents

\clearpage

\section[Supplementary Note 1 \\Wind data]{Supplementary Note 1 \\Wind data}
\label{sec: wind}
%

We installed a 2~m high wind tower in the center of our new experimental site and collected the wind data of
the local airport located 10~km east. We use the local tower for the measurements of the transport threshold
and the saturation length (see Figs.~2 and 3 of the main manuscript). We check the consistency between the
local wind data and those collected at the airport. For the long-term experiment, we only use the wind data
of the local airport.

\begin{figure}
\centerline{
\includegraphics[width=0.88\linewidth]{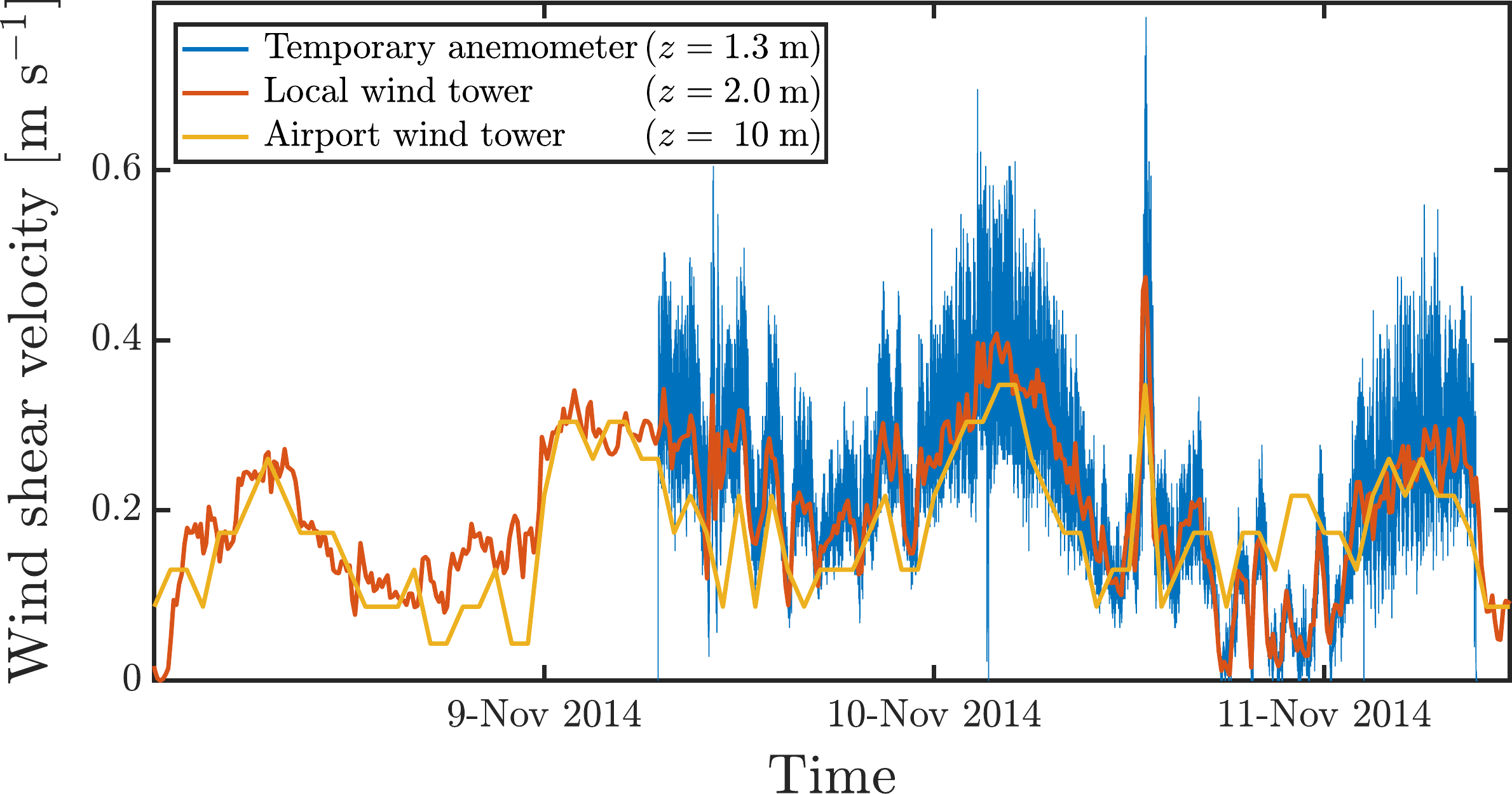}
}
\caption{{\bf Comparison of wind speed data.}
Wind shear velocity derived from temporary wind speed measurements made during the threshold
experiment (blue) and simultaneous measurements obtained on the local (red) and the airport
wind towers (yellow). There is a general agreement between all these data sets.
}
\label{fig: compa_wind}
\end{figure}

Fig.~\ref{fig: compa_wind} shows the shear velocity derived from a local wind measurement, the 2~m high wind
tower and the airport meteorological tower during the transport threshold experiment (see Fig.~2 of the main
manuscript). All these data are consistent with each other. As a consequence, the threshold shear stress
derived from the local measurement can be extrapolated to other wind data to compute sand fluxes
(Sec.~\ref{sec: flux}).


\section[Supplementary Note 2 \\Grain size distribution]{Supplementary Note 2 \\Grain size distribution}
\label{sec: d}
\begin{SCfigure}[][h]
\includegraphics[width=0.6\linewidth]{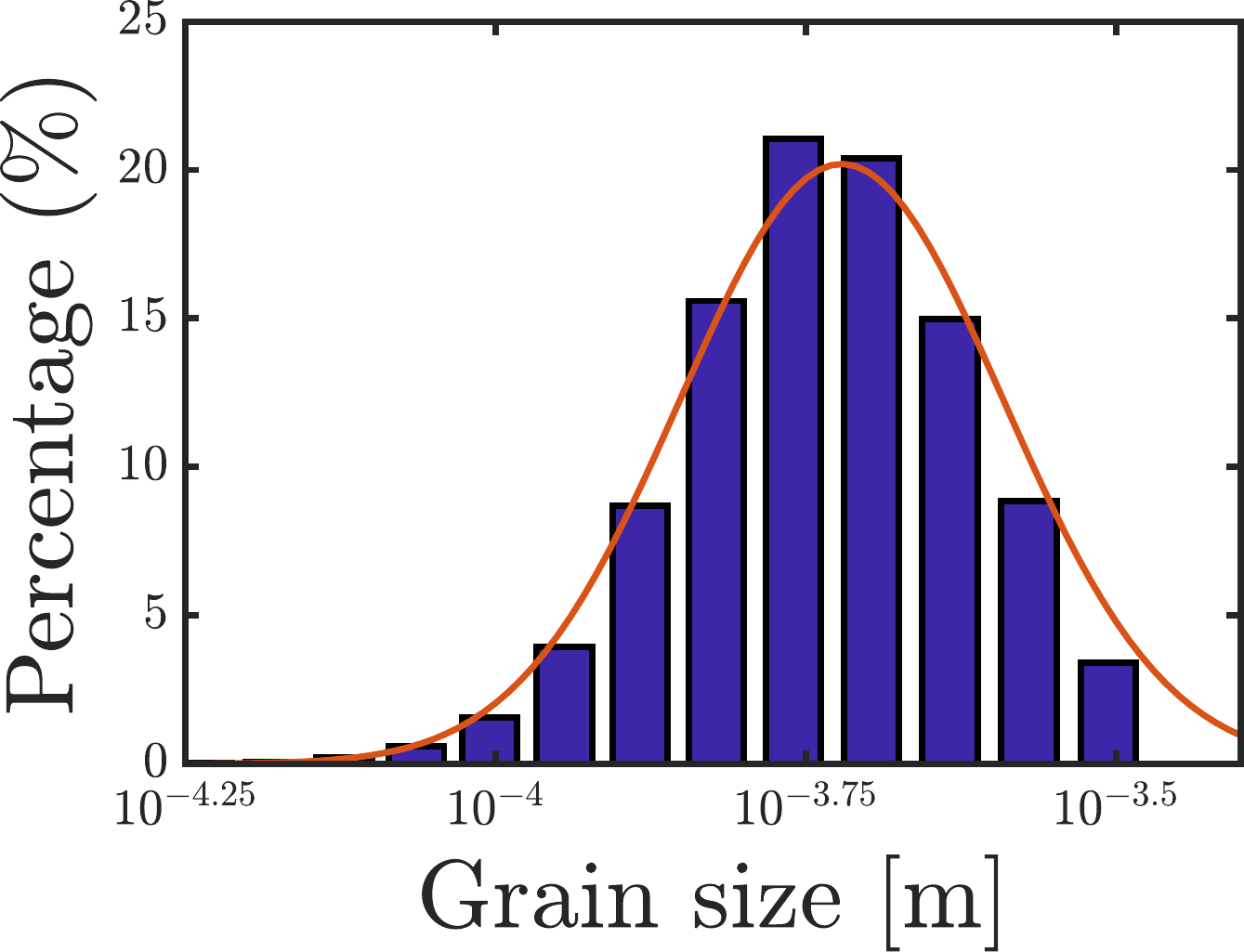}
\caption{{\bf Grain size distribution in the Tengger Desert.}
The line is the best fit using a lognormal distribution with a mean
value $d=190\; \mu {\rm m} \approx 10^{-3.72} \; {\rm m}$.
}
\label{fig: grain}
\end{SCfigure}

The landscape-scale experiment site is located close to the oasis city of Shapotu
at 8~km from the Yellow River in the Tengger Desert, which covers
an area of about 36,700 km$^2$ in the northwest part of the Zhongwei County
in the Ningxia Hui Autonomous Region of the People's Republic of China
(37\textdegree{}~31\textasciiacute{}~N, 105\textdegree{}~E). This desert is characterized
by a lognormal grain size distribution with a mean value $d$=190$~\mu$m
(Fig.~\ref{fig: grain}).

\begin{SCtable}[][t]
\begin{tabular}{lcc}
Variable & Units & Value \\
\hline
\hline
Acceleration of gravity $g$ &${\rm m\,s}^{-2}$ &9.81\\
Grain size $d$ & m & 190$\times 10^{-6}$\\
Air density $\rho_{\rm f}$ &${\rm kg\,m}^{-3}$ & 1.29\\
Grain density $\rho_{\rm s}$ &${\rm kg\,m}^{-3}$ & 2.55$\times 10^3$\\
Aerodynamic roughness $z_{\rm s}$ & m & 10$^{-3}$\\
von-K\'arm\'an constant $\kappa$ & $\emptyset$ & 0.4\\
\hline
\multicolumn{3}{c}{Shear velocity and sand flux on a flat sand bed}\\
\hline
Threshold shear velocity $u_{\rm th}\;\;\;\;\;\;\;\;\;\;\;$ & m$\;$s$^{-1}$ & 0.19 \\
Mean shear velocity $\langle u_* \rangle$ & m$\;$s$^{-1}$ & 0.29 \\
$\langle u_* \rangle/u_{\rm th}$ & $\emptyset$ & 1.5 \\
${\rm DP}=\langle\|\overrightarrow{Q}\|\rangle$ & m$^2\;$yr$^{-1}$ & 18.4 \\
${\rm RDP}=\|\langle\overrightarrow{Q}\rangle\|$ & m$^2\;$yr$^{-1}$ & \phantom{0}5.7 \\
RDP/DP & $\emptyset$ & 0.32 \\
${\rm RDD}$& mod 360\textdegree & 266.6 \\
\hline
\hline
\end{tabular}
$\;\;\;\;\;$
\caption{
{\bf Shear velocity and sand fluxes derived from the airport wind data from January 1, 2013
to October 31, 2017}. The wind and sand flux roses are shown in Fig.~1B of the main manuscript.
See text and Eqs.~\ref{eq: u1}-\ref{eq: mu2} for the description of all the variables. The
resultant drift direction (${\rm RDD}$) is measured counterclockwise from East.
}
\label{tab: 1}
\end{SCtable}

\section[Supplementary Note 3\\ Transport properties on a flat sand bed]{Supplementary Note 3\\ Transport properties on a flat sand bed}
\label{sec: flux}
Wind data are used to predict sand flux properties on a flat sand bed. Tab.~\ref{tab: 1}
shows the results obtained using the formalism that follows.

Wind measurements provide the wind speed  $u_i$ and direction $\vec{x}_i$ at different times
$t_i,\; i\in[1; N]$. For each  time step $i$, the shear velocity writes
\begin{equation}
u_*^i=\displaystyle \frac{u_i\kappa}{\log(z/z_s)},
\label{eq: u1}
\end{equation}
where $z$ is the height at which the wind velocity $u_i$ has been measured and $\kappa$ the
von-K\'arm\'an constant. Instead of the geometric roughness that depends only on grain size,
we consider here the aerodynamic roughness $z_{\rm s}$ that accounts for the height of the
transport layer in which saltating grains modify the vertical wind velocity profile. The
value of the threshold shear velocity for motion inception is determined using the formula
calibrated by {\em Iversen and Rasmussen}\cite{iversen1999effect}
\begin{equation}
u_{\rm th}=0.1 \displaystyle \sqrt{\frac{\rho_{\rm s}}{\rho_{\rm f}}gd}.
\label{eq: u2}
\end{equation}
Using the gravitational acceleration $g$, the grain to fluid density ratio
$\rho_{\rm s}/\rho_{\rm f}\simeq 2.05 \times 10^3$ and the grain diameter $d=190\,\mu {\rm m}$, we
find $\rm u_{\rm th}=0.19\, m \,s^{-1}$, which corresponds to a threshold wind speed of
$u_{10}=4.4\, {\rm m} \,{\rm s}^{-1}$ ten meters above the ground. It is close to the values we measured in the
field, which are $u_{\rm th}=0.23\pm 0.0.4$ and $u_{10}=5.3\pm 0.92 \, {\rm m} \,{\rm s}^{-1}$.

For each time step $i$, the
saturated sand flux $\overrightarrow{Q_i}$ on a flat sand bed is computed from the
relationship proposed by {\em Ungar and Haff}\cite{Unga87} and calibrated by
{\em Dur\'an et al.}\cite{Dura11}
\begin{equation}
Q_{\rm sat}(u_*)=
\begin{cases}
25\,\displaystyle \frac{\rho_{\rm f}}{\rho_{\rm s}} \sqrt{\displaystyle \frac{d}{g}}\left(u_*^2-u_{\rm th}^2 \right)
& \text{{\small for}  $u_* > u_{\rm th}$},\\
0 & \text{{\small else}}.
\end{cases}
\label{eq: sed_flux}
\end{equation}
In this formula, the prefactor takes into account a dune compactness of $0.6$.

From the individual saturated sand flux vectors $\overrightarrow{Q_i}$, we estimate the
mean sand flux vector on a flat erodible bed
\begin{equation}
\langle\overrightarrow{Q}\rangle=\displaystyle
{\displaystyle \sum_{i=2}^{N}  \overrightarrow{Q_i}  \delta t_i}\bigg/
{\displaystyle \sum_{i=2}^{N} \delta t_i},
\label{eq: mq}
\end{equation}
where $\delta t_i=t_i-t_{i-1}.$ The norm of the mean sand flux is usually called the
resultant drift potential:
\begin{equation}
{\rm RDP}=\|\langle\overrightarrow{Q}\rangle\|.
\label{eq: rdp}
\end{equation}
This quantity is highly dependent on the wind regime. Since it is a vectorial sum,
the contributions of winds from opposite directions cancel each other out. For
the entire time period, we also calculate the drift potential,
\begin{equation}
{\rm DP}=\displaystyle
{\displaystyle  \sum_{i=2}^{N} \left\| \overrightarrow{Q_i} \right\| \delta t_i }\bigg/
{\displaystyle \sum_{i=2}^{N} \delta t_i}.
\label{eq: dp}
\end{equation}
Unlike the resultant drift potential, this mean sand flux does not take into account the
orientation of the individual sand fluxes computed from the successive wind
measurements\cite{fryberger1979dune}.

The ratio $\rm RDP/DP$ is a non-dimensional parameter, which is often used to characterize
the directional variability of the wind regimes\cite{pearce2005frequency,tsoar2005sand}:
$\rm RDP/DP$ $\to 1$ indicates that sediment transport tends to be unidirectional;
$\rm RDP/DP$ $\to 0$ indicates that most of the transport components cancel each other.
Finally, RDD is the resultant drift direction, i.e., the direction of
$\langle\overrightarrow{Q}\rangle$.

The mean shear velocity $\langle u_* \rangle$ is defined as the shear velocity averaged
over the transport periods. i.e. when $Q_{\rm sat}>0$. Using the Heaviside function
$H_{u_{\rm th}}$ defined as
\begin{equation}
H_{u_{\rm th}}=
\begin{cases}
1 & \text{{\small for}  $u_* > u_{\rm th}$},\\
0 & \text{{\small else}}.
\end{cases}
\label{eq: heavi}
\end{equation}
the mean shear velocity can be defined directly from the shear velocity
\begin{equation}
\langle u_* \rangle = {\displaystyle \sum_{i=2}^{N} H_{u_{\rm th}}^i u_*^i} \bigg/ {\displaystyle \sum_{i=2}^{N} H_{u_{\rm th}}^i},
\label{eq: mu1}
\end{equation}
or from the integrated flux using the inverse function $Q_{\rm sat}^{-1}$ of the transport
law (Eq.~\ref{eq: sed_flux})
\begin{equation}
\langle u_* \rangle = Q_{\rm sat}^{-1} \left( {\rm DP}\times {\displaystyle \sum_{i=2}^{N} \delta t_i} \bigg/ {\displaystyle \sum_{i=2}^{N} H_{u_{\rm th}}^i \delta t_i} \right).
\label{eq: mu2}
\end{equation}
These two estimations of $\langle u_* \rangle$ are close to each other considering wind
data from the Tengger Desert.


\section[Supplementary Note 4\\ Estimating the saturation length $l_{\rm sat}$ in the field]{Supplementary Note 4\\ Estimating the saturation length $l_{\rm sat}$ in the field}
\label{sec: lsat}
\begin{SCfigure}[1][!hbtp]
\centering
\includegraphics[width=0.55\linewidth]{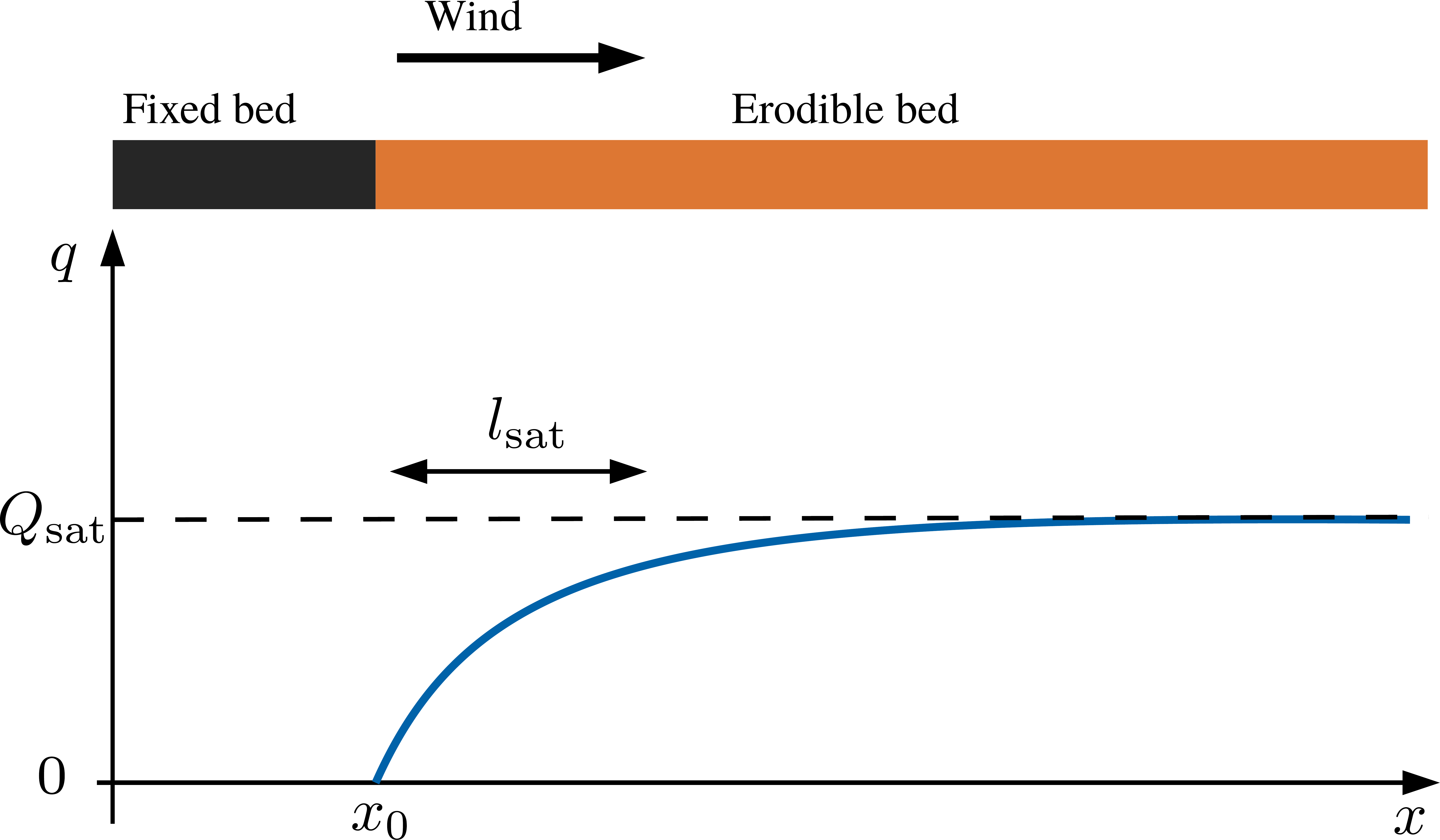}
\caption{
{\bf A schematic representation of the saturation length experiment}.
The saturation length may be described as the relaxation length
of the sand flux $q$ toward its saturated value $Q_{\rm sat}$. The experiment set-up is a flat
sand bed preceded by a flat non-erodible bed. When the wind blowing over this rigid surface reaches the
erodible bed, the sand flux relaxes exponentially towards its saturation value $Q_{\rm sat}$ over a
characteristic distance $l_{\rm sat}$ in the direction of the flow.
}
\label{fig: lsat}
\end{SCfigure}

Let us consider an infinite flat granular bed under a unidirectional wind in a statistically steady state.
Eventually, the transport rate reaches an equilibrium state due to the negative feedback between the
density of moving grains and the strength of the flow, which determines the saturated flux
$Q_{\rm sat}$ (Eq.~\ref{eq: sed_flux}). We consider now a situation for which the flow and the sediment
flux is non-homogeneous or unsteady in space or time. The actual flux $q$ does not immediately adjust to
the local value of the shear stress\cite{bBagn,Saue01,Nart09}. It needs some space and time to reach its
equilibrium $Q_{\rm sat}$-value (Fig.~\ref{fig: lsat}). Over bedforms, the transport is never far from its
saturated state, so it can be expressed by a first-order linear relaxation in both space and time
\begin{equation}
t_{\rm sat}\,\dfrac{\partial q }{\partial t}+l_{\rm sat}\,\dfrac{\partial q }{\partial x}=q_{\rm sat}-q,
\label{eq: lsat}
\end{equation}
where $l_{\rm sat}$ and $t_{\rm sat}$ are the saturation length and the saturation time, respectively\cite{Saue01,Andr02,Char06}.
The $t_{\rm sat}$-value is usually much smaller ($\approx 1$~s) than the characteristic time scale for
the evolution of the bed ($\approx 10^5$~s). Such a separation in scale justifies the simplifying assumption
that the fluid flow and sediment transport can be considered and computed as if the bed was fixed\cite{Dura11}.
Neglecting $t_{\rm sat}$, the saturation transient described by Eq.~\ref{eq: lsat} has been successfully applied to the description of
dune formation\cite{Nart09,Char06,Andr02,Andr05,Four10}. Under the configuration shown in Fig.~\ref{fig: lsat},
we have
\begin{equation}
q(x)=Q_{\rm sat}\left(1-\exp\left(\dfrac{x-x_0}{l_{\rm sat}}\right) \right).
\label{eq: lsat1}
\end{equation}
In practice, the saturation length $l_{\rm sat}$ scales as $(\rho_{\rm s}/\rho_{\rm f})d$ the distance needed
for one grain to be accelerated up to the wind velocity\cite{Andr02,Hers02,Andr04,Andr10}:
\begin{equation}
l_{\rm sat} \approx 2.2 \, \dfrac{\rho_{\rm s}}{\rho_{\rm f}}d,
\label{eq: ldrag}
\end{equation}
%
%

\begin{figure}
\centerline{
\includegraphics[width=1.00\linewidth]{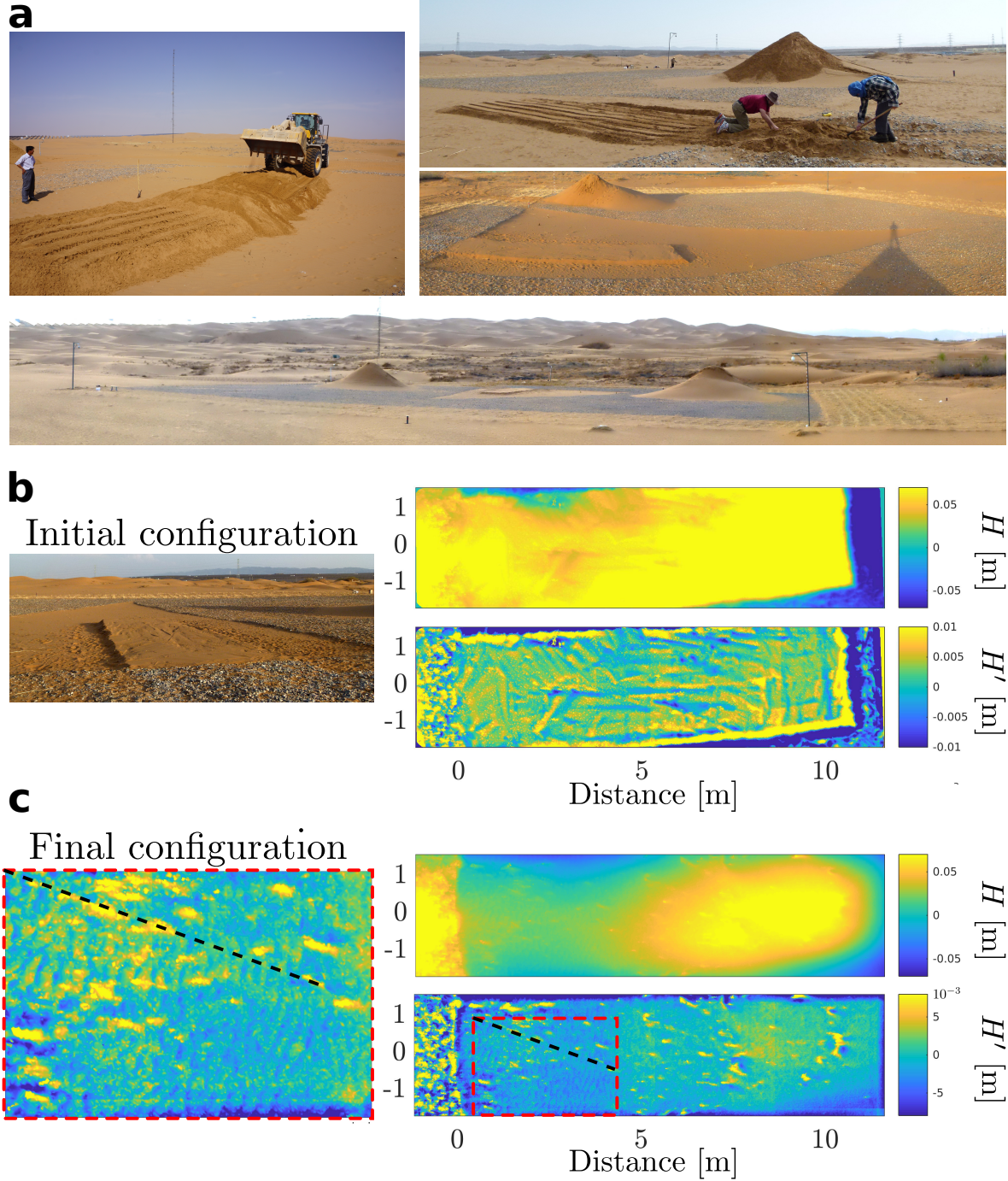}
}
\caption{
{\bf Measuring the saturation length $l_{\rm sat}$ in the field}.
{\bf (a)}
Setting up the $l_{\rm sat}$ experiment between the two sand piles of the elongating dune experiment
in a triangular area of bare sand within a flat bed armored with gravels. In this area, a rectangular
sand pile 12~m long and 3~m wide is placed downstream of the non-erodible gravel bed. The long axis
is aligned along the northwest-southeast direction, the orientation of the prevailing wind. The sand
bed is flattened so that there is no vertical step between the erodible and non-erodible zones.
{\bf (b)}
Elevation of the rectangular sand pile on April 22, 2014 before aeolian transport.
{\bf (c)}
Elevation of the rectangular sand pile on April 25, 2014 after the passage of a northwestern depression
(see flux rose in Fig.~3C of the main manuscript).
$H$ is the absolute elevation from a base level.
$H'$ is the difference of elevation with a smoothed surface obtained using a sliding window with a
radius of 20~cm. The inset in {\bf c} show the transect along which the elevation profile used to
estimate $l_{\rm sat}$ has been extracted (see Fig.~3B of the main manuscript). It is parallel to the
sand deposits that form downwind of topographic obstacles and perpendicular to aeolian ripples.
}
\label{fig: l_sat1}
\end{figure}

As shown in Fig.~\ref{fig: l_sat1} and in Fig.~3 of the main manuscript, we determine the value of
$l_{\rm sat}$ from the evolution of the topography of a rectangular sand pile placed downstream of
a non-erodible bed composed of gravels. We thus have the experimental setup that best reflects the
theoretical conditions presented in Fig.~\ref{fig: lsat}. The rectangular sand pile has a length of
12~m and a width of 3~m with its main axis aligned in the northwest-southeast direction, parallel to
the orientation of the prevailing wind (Fig.~\ref{fig: l_sat1}a). The initial sand pile was prepared
and scanned on April 22, 2014 (Fig.~\ref{fig: l_sat1}b). A storm occurred on April 24 with winds from
the north-northwest and irregular wind speeds reaching 15~m$\;$s$^{-1}$ at a height of 2~m. The sand
pile was scanned again on April 25 (Fig.~\ref{fig: l_sat1}c).

Considering the ideal assumption of our initially flat surface under a steady wind of constant
direction, strength and transport rate $q$, Eq.~\ref{eq: lsat1} can be combined with the
equation of conservation of mass to estimate the erosion rate
\begin{equation}
\dfrac{\partial h(x)}{\partial t}
= - \dfrac{\partial q(x)}{\partial x}
= - \dfrac{Q_{\rm sat}}{l_{\rm sat}} \exp\left(\dfrac{x-x_0}{l_{\rm sat}}\right).
\label{eq: mass}
\end{equation}
This equation can be integrated to estimate the net erosion over a given time period. In this case
the exponential regime is expected to hold. However, it cannot be observed from the start of the
erodible bed for several reasons, both theoretical and experimental. The main reasons are related
to (1) the natural variability of wind speed and direction and (2) the development of a
discontinuity in the topographic profile between the non-erodible and the erodible beds. In
addition, Eq.~\ref{eq: mass} neither takes into account the possible dependence of $l_{\rm sat}$
on wind speed\cite{Paht13}, nor the spatial shift associated with the establishment of a transport
layer when the sand flux starts from zero\cite{Andr10}.

Using the field data and despite the number of simplifying assumptions, we study the difference
in height $\Delta H$ between the two surface elevations to look for zones where
\begin{equation}
\Delta H \sim - \exp(x/l_{\rm sat}).
\label{eq: decay}
\end{equation}
In practice, we use 2D elevation profiles aligned with the primary sand transport
direction  during the storm. It forms an angle of 22\textdegree $\;$with the orientation of the
initial sand pile. This angle is estimated from the sand flux rose as well as from the
orientation of ripples and accretion mounds (Fig.~\ref{fig: l_sat1}c).

The slope of the exponential decay in Eq.~\ref{eq: decay} gives the value of $l_{\rm sat}$ (Fig.~3D
of the main manuscript). We find a value of 0.95~m from our field data, a value that can be directly
compared to the 0.83~m predicted by Eq.~\ref{eq: ldrag} and the parameters given in Tab.~\ref{tab: 1}.

%

\section[Supplementary Note 5\\ Upwind velocity shift at dune crests and troughs]{Supplementary Note 5\\ Upwind velocity shift at dune crests and troughs}
\label{sec: shift}
%
%
\begin{figure}
\centerline{
\includegraphics[width=.94\linewidth]{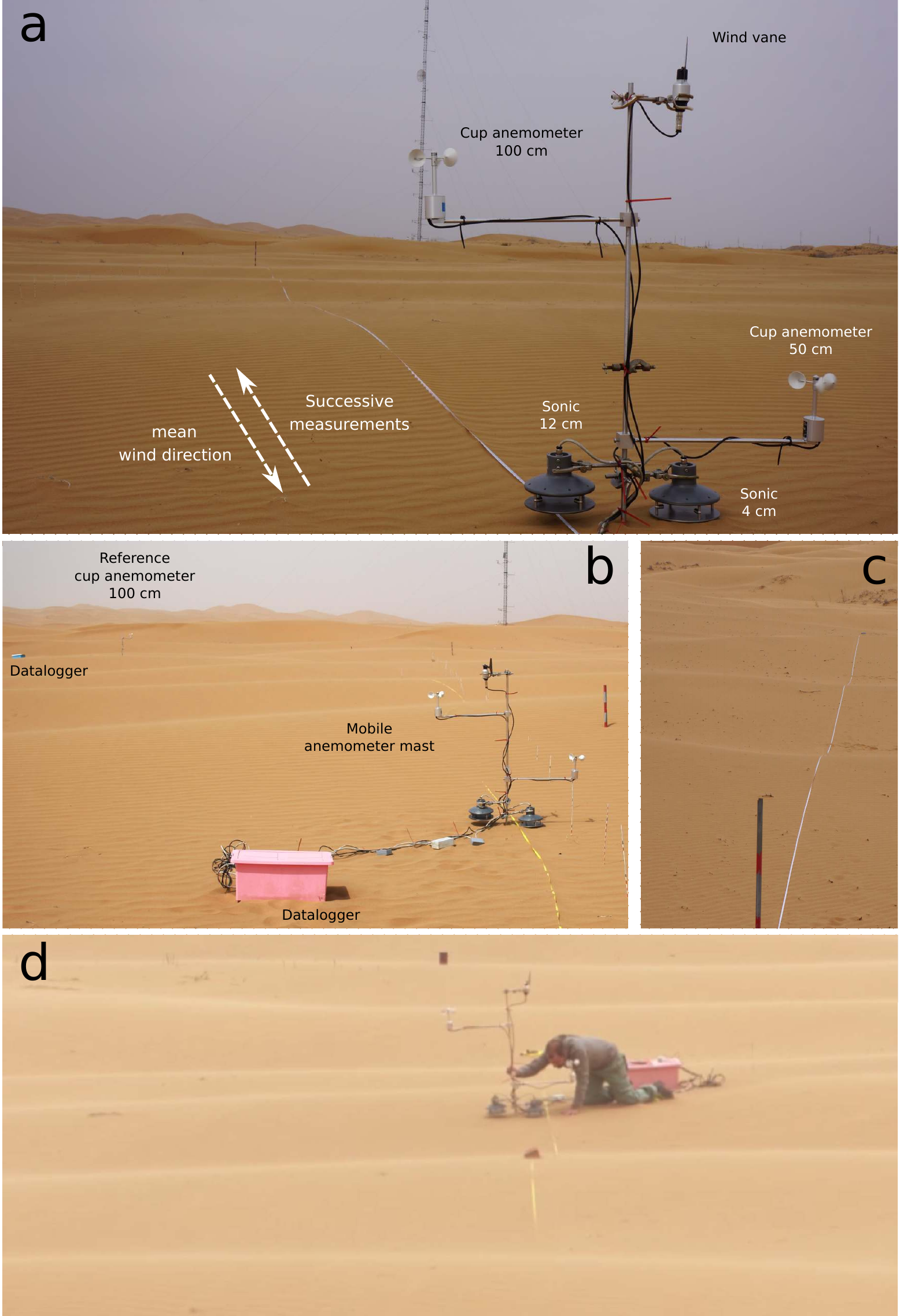}
}
\caption{
{\bf Experimental set-up to estimate the upwind velocity shift at the crests and troughs of incipient dunes.}
{\bf (a)}
The mobile anemometer mast, with anemometers located at heights of 4 and 12~cm in the inner layer and
at heights of 50 and 100~cm in the outer layer. Arrows show the mean wind direction and the
measurement path.
{\bf (b)} The anemometer mast and the reference anemometer.
{\bf (c)} A measurement path perpendicular to the crest orientation of the bed instability.
{\bf (d)} Setting a new measurement for 10~min with a sampling frequency of 1~Hz.
}
\label{fig: mast}
\end{figure}

\subsection{The inner and outer layers}
\label{sec: tshift}
Flows that are topographically forced by obstacles (e.g., hills or sand dunes) accelerates on the upwind
slopes and decelerates on the downwind slopes. Then, in order to study sediment transport over a dune, we
should first describe some properties of the turbulent flow over an undulating topography. Conceptually, as
proposed by {\em Jackson and Hunt (1975)}\cite{Jack75} in the limit of small amplitudes bedforms, the turbulent flow
over a sinuous bed elevation profile of wavelength $\lambda$ can be decomposed into two layers:
\begin{itemize}
\item
The outer layer is the external region (supposed infinite) where the pressure gradient set up by the topography
is balanced by the inertial forces. The streamlines follow the topography. At a given height, the wind speed is
maximum (minimum) above the top (bottom) of the topography. The amplitude of the perturbation vanishes on a
characteristic height that varies according to the wavelength of the topography. Then, sufficiently far above the
obstacle, the wind speed is finally equal to the undisturbed wind speed in the absence of topography (see
Eq.~\ref{eq: u1}).
\item
The inner layer is a zone in which the longitudinal pressure gradient exerted by the fluid is compensated
by the Reynolds shear stress induced by the turbulent motions at the surface of the bed. Then, the pressure
gradient is in phase quadrature with the topography and is maximum where the stoss slope is steepest. Hence,
there is an upwind velocity shift within the inner layer. The characteristic thickness $l$ of the inner layer
depends on both the wavelength of the bed elevation profile and on the aerodynamic roughness $z_{\rm s}$ so
that\cite{Jack75,Syke80}
\begin{equation}
\dfrac{l}{\lambda} \log^2 \left( \dfrac{l}{z_s} \right) = 2 \kappa^2.
\label{eq: inner}
\end{equation}
For typical value of the aerodynamic roughness ($z_{\rm s}<10^{-2}$~m) and wavelengths of hundreds of
meters, the inner layer is always confined in a meter scale envelop above the topography. For
$\lambda=20\;{\rm m}$ and $z_{\rm s}=10^{-3}\;{\rm m}$, common values observed during the development of
aeolian bedforms under high wind speed, the inner layer is less than 20~cm.
\end{itemize}

\subsection{Measuring the upwind velocity shift within the inner layer}
\label{sec: mshift}
A sufficiently strong wind and a stable wind orientation are essential for the quality of the measurements.
The sequence of measurements consists in moving an anemoter mast upwind along a given profile of known elevation.
The entire procedure to estimate the upwind velocity shift on dunes is described in full details in
{\em Claudin et al. (2013)}\cite{Clau13}. We follow here the same procedure with the specifications detailed below.

We performed our measurements on low dunes with sinusoidal shape. The density of measurements is high near the crests and
troughs ($\approx 20\;{\rm cm}$) but lower in the steeper sections ($\approx 1\;{\rm m}$). Each measurement lasts
10~min with a sampling frequency of 1~Hz. The mobile anemometer mast has sonic anemometers located at heights of 4
and 12~cm and cup anemometers at heights of 50 and 100~cm (Fig.~\ref{fig: mast}a). Given the natural variability in
wind speed between two measurements, the wind speeds measured along the mobile mast are normalized by the wind speed
measured at one meter high by a reference cup anemometer located at the top of a larger dune in the vicinity
(Fig.~\ref{fig: mast}b). The normalized wind speeds along the profile can then reveal how sensitive is the flow to
topography at different heights. As shown in Fig.~4B of the main manuscript, in the limit of small sinusoidal
oscillations, the wind profiles at all heights reflect the topography, i.e., the spatial variation of the wind
speed exhibits the same wavelength as the bedforms. Most importantly for our present purposes, the flow is in phase
with the topography at heights of 50 and 100~cm and in phase advance at heights of 4 and 12~cm. It indicates that
the inner layer has a thickness between 12 and 50~cm, a range of value that covers entirely the values given by
Eq.~\ref{eq: inner} for aeolian dune systems on Earth.

%
\subsection{From the upwind velocity shift to the hydrodynamic parameters $A$ and $B$}
\label{sec: ab}
On several occasions during our field campaigns, we have measured wind velocity in the inner and outer layers
on a succession of crests and troughs to estimate the phase shift over more than a wavelength of the dune
pattern (Fig.~\ref{fig: mast}c,d). Here, we focus only on individual dune crests or troughs, which are
approximated by
\begin{equation}
H(x) = H_{\rm ref}+ h \cos(kx),
\label{Zprofile}
\end{equation}
where $k$ is the wave number (wavelength $\lambda = 2\pi/k$), $h$ the amplitude. The position of the dune
crests and troughs are set at $x=0$ and $x=\lambda/2$, respectively. The arbitrary reference level $H_{\rm ref}$
is here chosen such that $H>0$. Fitting Eq.~\ref{Zprofile} to the dune elevation data gives the values of $k$
and $h$. The dune aspect ratio is $R=2h/\lambda=h k/\pi$. When the aspect ratio is below $\simeq 0.05$, we expect
that the (low) perturbation calculation of the aerodynamics to be valid. A more refined analysis would involve the
computation of the Fourier transform of the dune profile, in order to account for a whole range of wave numbers
$k$.

Alike topography, the wind profile at a given height along the dune can also be fitted by a sinusoidal function
of the same wave number $k$ as for the dune elevation:
\begin{equation}
u_{\rm b}(x) = u_{\rm b}^0+ \delta u_{\rm b} \cos(kx+\varphi_{\rm b}).
\label{uprofile}
\end{equation}
As said in Sec.~\ref{sec: tshift}, the topography $Z(x)$ and the wind velocity $u_{\rm b}(x)$ are not in phase
for anemometers at heights of 4 and 12~cm: the velocity reaches its maximum upstream of the crest. Here the
phase difference is $\varphi_{\rm b}$ so that the upwind velocity shift in the inner layer is
$\varphi_{\rm b}/k$.The two other fitting parameters are $u_{\rm b}^0$ and $\delta u_{\rm b}$.

Because the logarithmic law of the wall (Eq.~\ref{eq: u1}) locally holds in the inner layer at each
position $x$, the velocity can be used as a proxy to calculate the basal shear stress with
$\tau_{\rm b} \propto \rho_{\rm f} u_{\rm b}^2$. By expansion to the first order, we can then write:
\begin{equation}
\tau_{\rm b}(x) \propto \rho_{\rm f} (u_{\rm b}^0)^2 \left( 1 + kh \left( A\cos(kx) - B\sin(kx) \right) \right),
\label{taubprofile}
\end{equation}
where $A$ and $B$ are given by
\begin{equation}
A = 2 \times \frac{\delta u_{\rm b}}{u_{\rm b}^0} \times \dfrac{\cos(\varphi_{\rm b})}{kh}
\;\;\;\;\;\;\;\;\;\;\;\;\;\;\; \mbox{and} \;\;\;\;\;\;\;\;\;\;\;\;\;\;\;
B = 2 \times \frac{\delta u_{\rm b}}{u_{\rm b}^0} \times \dfrac{\sin(\varphi_{\rm b})}{kh} \, .
\label{AandBfromVel}
\end{equation}
Eqs.~\ref{taubprofile}  and \ref{AandBfromVel} are specific to the quadratic relation between $\tau_{\rm b}$
to $u_{\rm b}$. Since sediment transport is controlled by the basal shear stress, these two parameters $A$
and $B$ are all that are needed as aerodynamical inputs in bedform evolution models. As described in the
introduction of the main manuscript, they are of fundamental importance for the understanding of the mechanisms
of dune growth and dune size-election (see also Sec.~\ref{sec: inst} and Eqs.~\ref{eq: disp1}-\ref{eq: disp2}).
Tab.~\ref{tab: tab3} shows the results obtained in April and November 2015 in our experiment. We observe no
significant trend in the variation of $A$ and $B$ during dune growth.


\begin{table}[h!]
\begin{center}
\begin{tabular}{clccp{11.5cm}}
\hline
\hline
{\bf Variable} && {\bf Units} && {\bf Description} \\
\hline
\multicolumn{5}{c}{\bf{Topography}} \\
$\lambda$ && m\phantom{$^{-1}$} && dune wavelength \\
$k$ && m$^{-1}$ && Dune wave number\\
$h$ && m\phantom{$^{-1}$} && Amplitude of the dune \\
$R$ && $\emptyset$ && Dune aspect ratio \\
\hline
\multicolumn{5}{c}{{\bf{Flow}}} \\
$\delta u_{\rm b}$ && ${\rm m\; s}^{-1}$ && Amplitude of wind speed variation in the inner layer.\\
$u_{\rm b}^0$ && ${\rm m\; s}^{-1}$ && Mean wind speed in the inner layer. \\
$\varphi_{\rm b}$ && $^{\rm o}$ && Phase shift between the flow and the topography in the inner layer.\\
\hline
\multicolumn{5}{c}{{\bf{Hydrodynamic parameters}}} \\
$A$ && $\emptyset$ && In-phase hydrodynamic parameter\\
$B$ && $\emptyset$ && In-quadrature hydrodynamic parameter\\
\hline
\hline
\end{tabular}
%

\begin{tabular}{cccccccccccp{6cm}}
& {\bf 14/04} & {\bf 15/04} & {\bf 15/04$^*$} & {\bf 15/04} & {\bf 16/04} & {\bf 18/04} & {\bf 03/11} & {\bf 03/11$^*$} & {\bf 03/11} & {\bf 15/11} \\
\hline
\multicolumn{11}{c}{{\bf{Topography}}} \\
$\lambda$ & 14.84 & 18.57 & 10.50 & 16.45 & 21.92 & 14.51 & 25.05 & 14.49 & 21.06 & 14.33 \\
$k$ & 0.423 & 0.338 & 0.598 & 0.382 & 0.287 & 0.433 & 0.251 & 0.437 & 0.298 & 0.439 \\
$h$ & 0.297 & 0.262 & 0.185 & 0.414 & 0.358 & 0.338 & 0.662 & 0.323 & 0.264 & 0.361 \\
$R$ & 0.020 & 0.014 & 0.018 & 0.025 & 0.016 & 0.023 & 0.026 & 0.022 & 0.012 & 0.025 \\
\hline
\multicolumn{11}{c}{{\bf{Flow}}} \\
$\delta u_{\rm b}$ & 0.126 & 0.111 & 0.108 & 0.198 & 0.0624  &0.143 & 0.306 & 0.134 & 0.132 & 0.174 \\
$u_{\rm b}^0$ & 0.683 & 0.650 & 0.656 & 0.678 & 0.743 & 0.701 & 0.568 & 0.622 & 0.739 & 0.661 \\
$\varphi_{\rm b}$ & 25.51 &  11.89 &  15.95 &  20.79 &  45.95 &  10.92 &  11.36 &  19.13 &  17.42 & 9.80 \\
\hline
\multicolumn{11}{c}{{\bf{Hydrodynamic parameters}}} \\
$A$ & 2.648 & 3.770 & 2.860 & 3.453 & 1.138 & 2.737 & 6.362 & 2.906 & 4.328 & 3.278 \\
$B$ & 1.264 & 0.794 & 0.817 & 1.311 & 1.176 & 0.528 & 1.278 & 1.008 & 1.358 & 0.566 \\
\hline
\hline
\end{tabular}\caption{{\bf Topography and wind measurements carried out in April and November 2015}.
Measurements on dune troughs are shown with a $^*$ on April 4 and November 3, 2015. They are located
between two dune crests along the same transect to estimate the phase shift over more than one wavelength
of the dune pattern.
}
\label{tab: tab3}
\end{center}
\end{table}


\section[Supplementary Note 6\\ Dispersion diagram from successive topographic surveys]{Supplementary Note 6\\ Dispersion diagram from successive topographic surveys}
\label{sec: inst}
Dispersion relations are plotted as growth rate $\sigma$ and phase velocity $c$ with respect to
wave number $k$. They are used in linear stability analysis to identify the most unstable wavelength,
which is likely to be observed. We describe here how we measure the growth rate of dunes as a function
of their wave numbers $k$ from successive topographic surveys throughout the duration of our
experiment. Before, we present the rational for studying dune formation as a linear instability in
a landscape scale experiment.

\subsection{Aeolian dune formation as a linear instability}
\label{sec: tinst}
In what follows, the term instability characterizes the growth of a small perturbation
in a system, which is often considered to be as homogeneous and simple as possible. Conversely,
stability refers to the ability of this system to return to its original state when perturbed.
The main objective of stability analysis is to identify the ranges of perturbation wavelengths
over which the system exhibits stable or unstable behavior. Quantitative approaches consists of
estimating the initial growth rate of each mode (wavelength) considering that all these modes
are independent of each other. Obviously, the highest and zero growth rates are particularly
important. The highest ones are associated with the most unstable modes, which will have the
greatest impact on the system. Those at zero are neutral modes that often mark the transition
between stable and unstable regimes.

The analysis of the time and length scales of instabilities by means of linearized equations is a
standard approach in hydrodynamics and many other branches of physics. By identifying stabilizing
and destabilizing mechanisms, these linear stability analysis reveal how they together govern the
the evolution of the system over the entire range of possible wavelengths. During the first stage
of the instability (the linear regime), the initial wavelength $\lambda_{\rm g}$ of the perturbation
is assumed to stay constant, whereas its amplitude $h$ grows exponentially with time
\begin{equation}
h(k_{\rm g},t) \propto \exp(\sigma(k_{\rm g}) t).
\label{eq: exp}
\end{equation}
In this expression, $k_{\rm g}=2\pi/\lambda_{\rm g}$ is the wave number and $\sigma(k_{\rm g})$ the
growth rate with units of frequency. Positive and negative growth rates correspond to unstable and
stable modes, respectively. The dispersion relation $\sigma(k)$ gives the growth rate value of the
perturbation as a function of the wave number $k$. The largest positive $\sigma$-value corresponds
to the most unstable mode $\{k_{\rm max},\, \lambda_{\rm max}\}$. Zero values corresponds to neutral
modes $\{k_0,\, \lambda_0\}$. In diffusive systems, dispersion relations are often characterized by
a transition from a stable to an unstable regime. for an increasing wavelength. In other words,
$\sigma(k)>0$ for $k<k_0$ and $\sigma(k)<0$ for $k>k_0$. Beyond the linear stage of the instability,
when the amplitude of the initial perturbation is too high, the dispersion relation no longer
applies because the different modes interact with one another. It is described as the non-linear
stage of the instability.

Here, we study the formation of aeolian dunes as a linear instability. More exactly, we focus on
the dependence and feedback between bed forms, wind flow and sand transport properties. The
governing equations lead to the theoretical expression of a dispersion relation as a function
of the different physical parameters of the system (see Eq.~1 of the main manuscript). Thus,
the minimum size for dunes is associated with a neutral mode and a transition from unstable to
stable regime for decreasing wavelength. The emergence and growth of periodic dune patterns are
associated with a most unstable mode which is going to prevail within the whole dune field. Based
on observation of mature aeolian dunes in nature, previous studies have measured the wavelength
and the migration rate of dunes in order to derive values of $\lambda_0$, $\lambda_{\rm max}$ and
$\sigma_{\rm max}$ under various conditions. Dispersion relations have been given less attention
or even disregarded, certainly because of the length and time scales involved in the mechanism of
aeolian dune growth. Then, a direct validation of the dune instability theory is to investigate
whether or not dispersion diagrams can be derived from field data. Another solution is to verify
whether the theoretical formalism used with the values of the underlying physical parameters
measured independently in the field is actually capable of accurately predicting the observations.

Here, we apply this methodology to the formation of dunes in a landscape scale experiment. Our
initial system is a sand bed after flattening by a bulldozer. In such an experimental set up, but
also in all natural dune environments, there are always heterogeneities and defects at all length
scales and our initial condition already contains all the perturbation wavelengths.

%
\subsection{The transition from the linear to the non-linear phases of dune growth}
\begin{figure}
\centerline{
\includegraphics[width=1.00\linewidth]{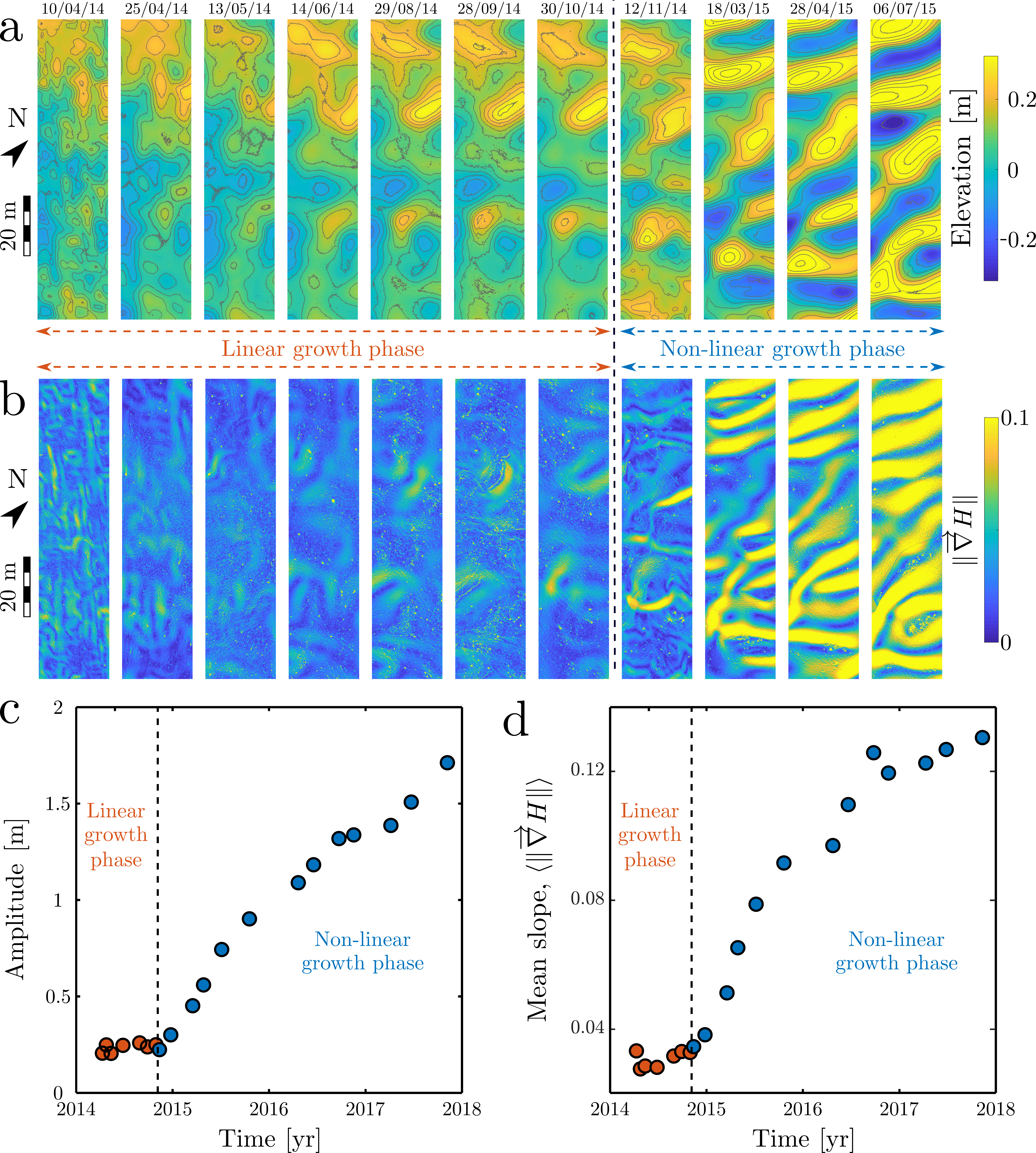}
}
\caption{
{\bf The transition from the linear to the non-linear phases of dune growth}.
{\bf (a)}
Elevation and
{\bf (b)}
slope maps during incipient dune growth from April 10, 2014 to July 6, 2015. The colormap in
Fig.~\ref{fig: id1}b is saturated to distinguish the slopes of bedforms in fall 2014, around
the transition from the linear to the non-linear phases of dune growth. In non-saturated areas,
ripples can be observed. A cable crosses the experimental field from 28 April 2015.
{\bf (c)}
The mean amplitude $2\sqrt{2}(\langle h^2\rangle-\langle h\rangle^2)^{1/2}$ of bedforms  and
{\bf (d)}
the mean slope $\langle \| \vec{\nabla} H \| \rangle$ with respect to time. The higher mean slope
value on April 10, 2014, after the flattening, is due to bulldozer tracks that can be seen in
Fig.\ref{fig: id1}b. Colors and dashed lines are used to separate the linear (orange) and the
non-linear (blue) phases of dune growth.  For comparison with slope data, Fig.~\ref{fig: id1}d
is the same as Fig.~5C of the main manuscript.
}
\label{fig: id1}
\end{figure}

\begin{figure}
\centerline{
\includegraphics[width=1.00\linewidth]{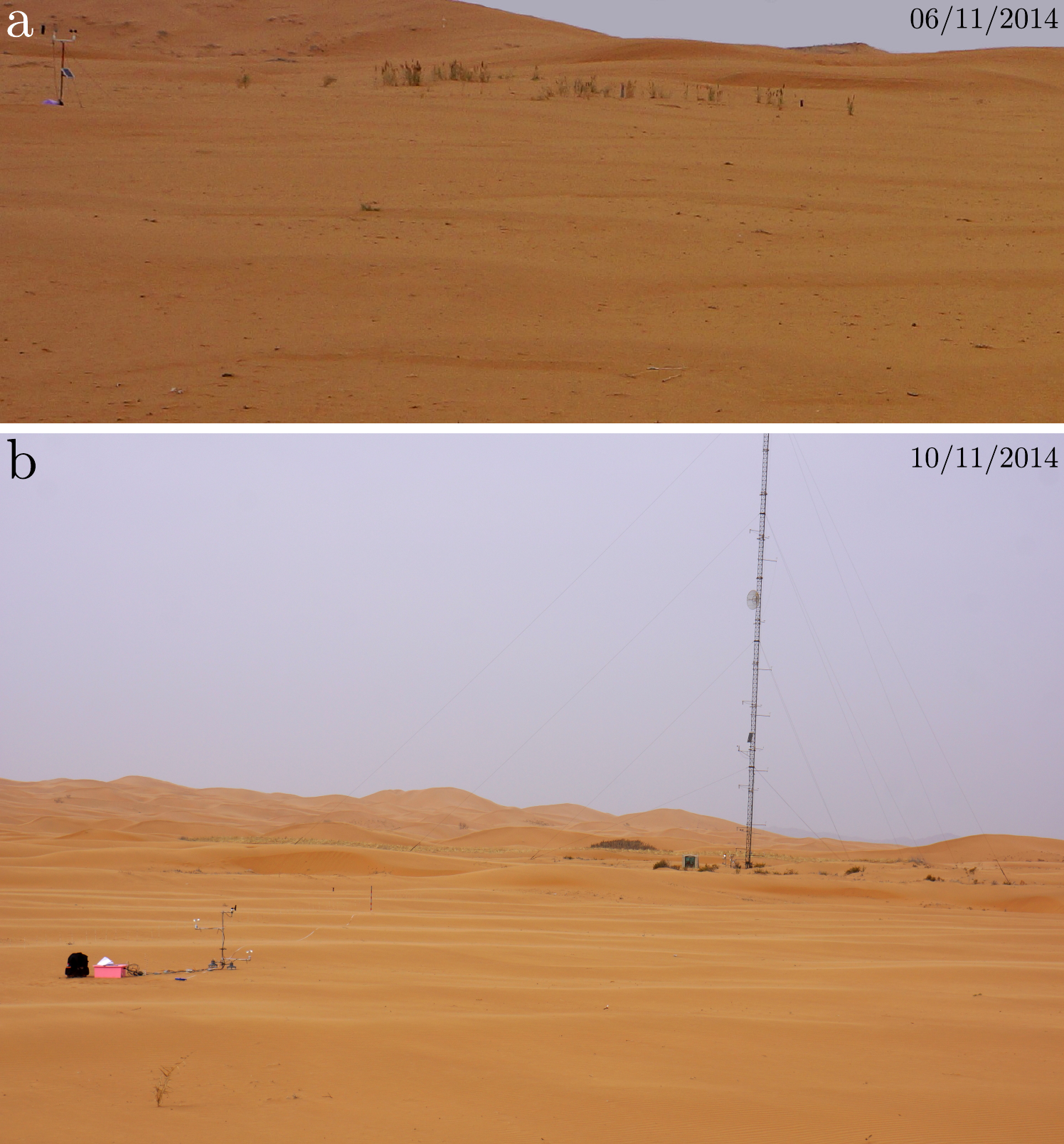}
}
\caption{
{\bf Bedforms during the transition from the linear to the non-linear phases of dune growth}.
{\bf (a)} Incipient slip faces a few centimetres high on November 6, 2014.
{\bf (b)} Periodic dune patterns on November 10, 2014.
This pictures can be compared to elevation and slope maps of October 30, 2014 and November 12, 2014
shown in Fig.~\ref{fig: id1}a and \ref{fig: id1}b.
}
\label{fig: id2}
\end{figure}

The continuous transition from the linear to the non-linear phases of dune growth is controlled
by dune aspect-ratio, which is the main control parameter for aerodynamic non-linearities.
Figs.~\ref{fig: id1}a-b show the elevation and slope maps during incipient dune growth in our
experimental field from April 10, 2014 to July 6, 2015. There is a significant change in slope
maps between October 30 and November 12, 2014. Over this time interval, the local slopes not only
become steeper, they also become spatially organized, just like dune crest, to form more regular
transverse structures across the flattened area. This occurs for a mean slope of about 0.03
(Fig.~\ref{fig: id1}c) at the same time as incipient slip faces a few centimeter high emerge
(Fig.~\ref{fig: id2}). The mean slope $\langle \| \vec{\nabla} H \| \rangle$ variations can be compared
to the amplitude of the bedforms in Figs.~\ref{fig: id1}b-c.  Based on these observations, we set
the transition between October 30 and November 12, 2014.  This transition is not spontaneous but
it is surely completed in April 2015, when the mean slope reaches a value of about 0.07.

%
%
%
%
%
\subsection{A transport time scale for dune growth under variable wind strength}
\label{sec: ta}
In order to quantify dune growth under variable wind strength, it is necessary to define a new time scale
that accounts for the intensity of transport. For example, periods during which there is no transport
should not be used as time increments, while storm periods should contribute more to the total time.

\begin{figure}[ht]
\centerline{
\includegraphics[width=0.8\linewidth]{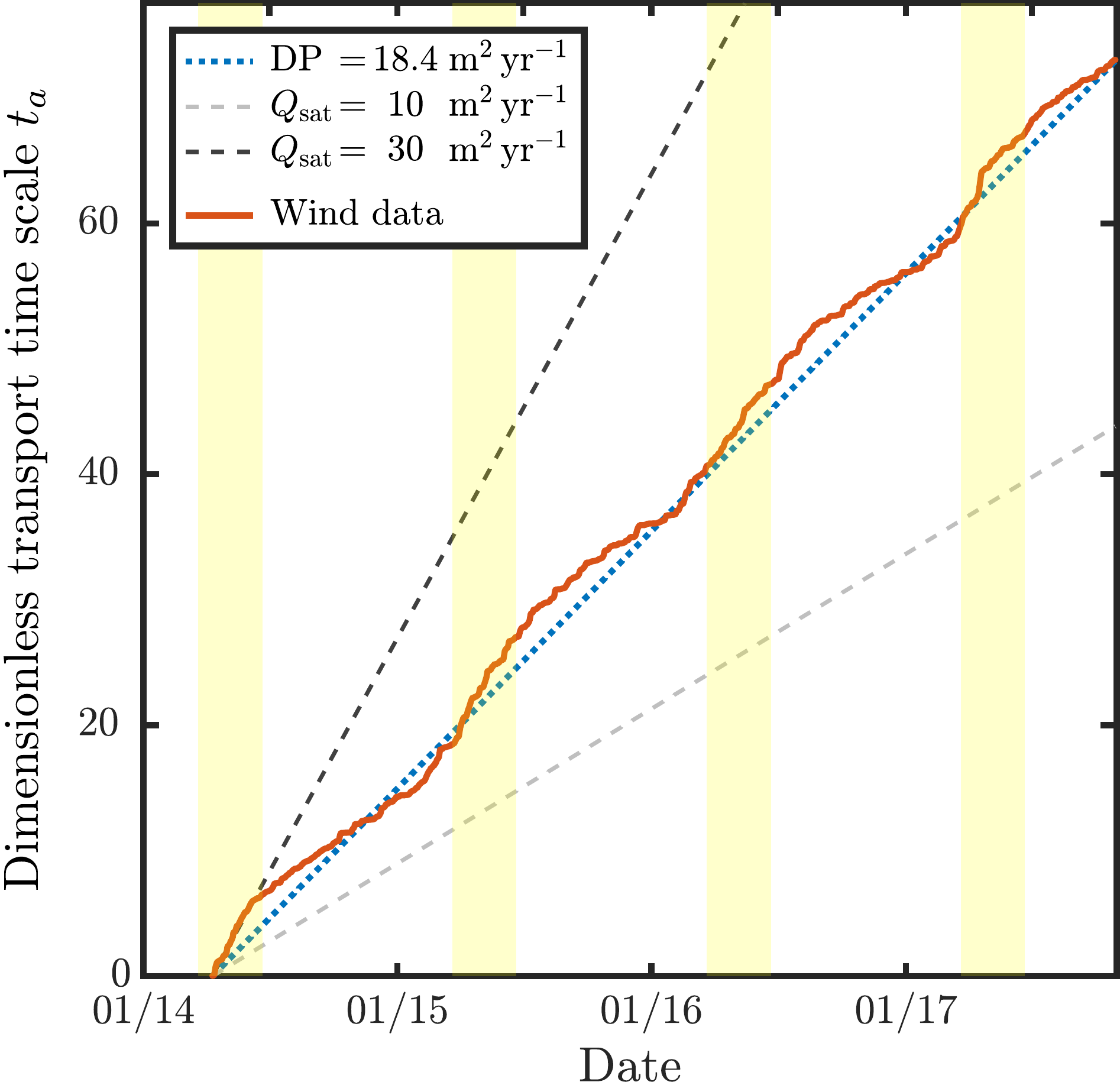}
}
\caption{
{\bf A dimensionless transport time scale for dune growth under variable wind strength}.
Using the saturated sand flux derived from wind data (Eq.~\ref{eq: sed_flux}), the red solid
line shows the dimensionless transport time scale $t_a$ as a function of time from April 10, 2014,
when the dune field was flattened. By definition, the mean slope is given by the RDP
(blue line, Eq.~\ref{eq: rdp}).
Light and dark gray lines show the transport time scale for constant sand fluxes of 10 and 30
${\rm m}^2\,{\rm yr}^{-1}$, respectively. Highlighted areas show the spring periods from March
21 to June 21.
}
\label{fig: ta}
\end{figure}

In practice, the new time scale is defined sequentially from the wind data using the saturated sand flux
(Eq.~\ref{eq: sed_flux}) and the saturation length $l_{\rm sat}$. Starting at $t=0$ at the flattening time,
the new times write
\begin{equation}
t_{\rm a}(t)=\dfrac{\displaystyle \sum_{i=1}^{t} \| \overrightarrow{Q_i} \| \delta t_i }{l_{\rm sat}^2}.
\end{equation}
This time scale is dimensionless allowing for comparison across different time series and various time
periods. Using the saturated sand flux derived from the local wind data (Eq.~\ref{eq: sed_flux}),
Fig.~\ref{fig: ta} shows the dimensionless times with respect to time from April 10, 2014 to October 31,
2017 using the saturation length $l_{\rm sat}=0.95~{\rm m}$ measured in the field (Fig.~3 of the main
manuscript). By definition, periods of stronger winds are associated with steeper slopes, and vice versa.
The choice of the characteristic length for the computation of the dimensionless transport time scale,
(here $l_{\rm sat}$) is of minor significance since we consider it to be constant.

%
%
%
\subsection{Topographic surveys during experimental dune growth}
\label{sec: flat}
\begin{figure}
\centerline{
\includegraphics[width=1.00\linewidth]{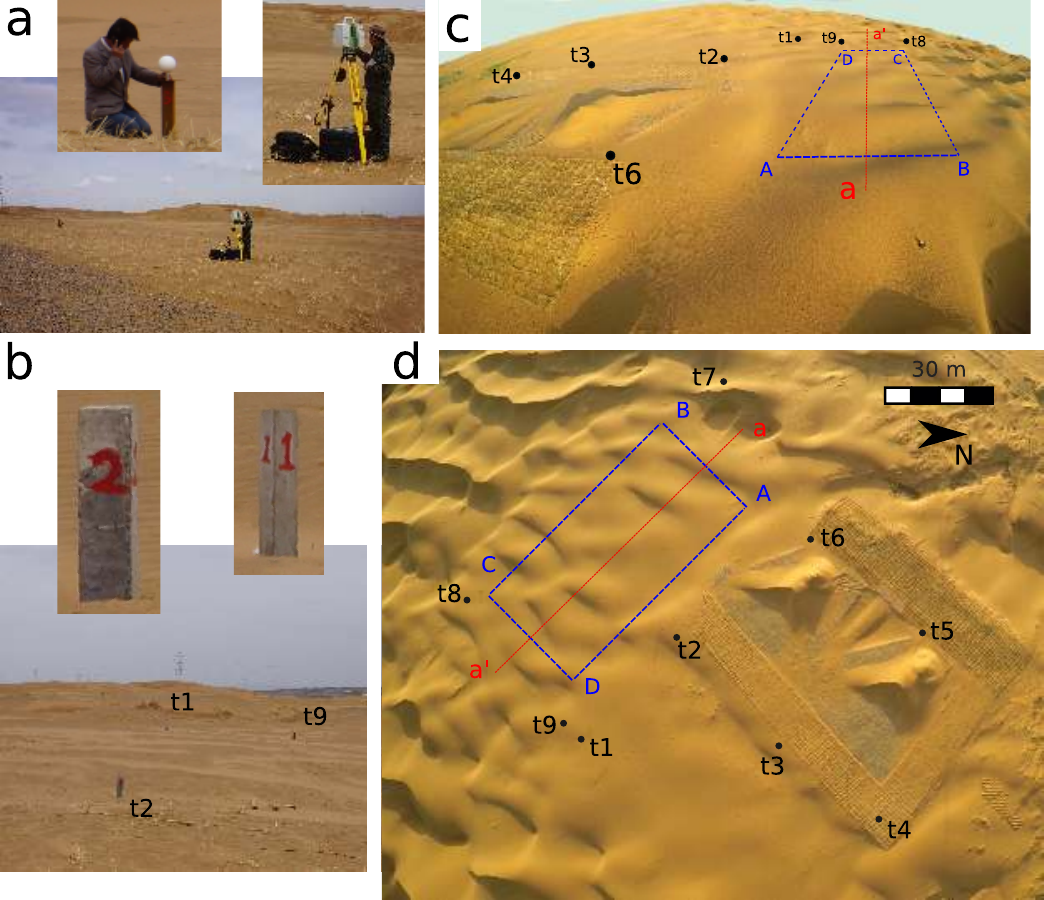}
}
\caption{
{\bf Topographic surveys during experimental dune growth}.
{\bf (a)} Ground laser scanning in the landscape-scale experiment.
{\bf (b)} Ground view of three posts of the local reference system.
{\bf (c,d)}
Reference system (dots) and selected area (blue rectangle) for the entire duration of the experiment.
All 2D elevation profiles are taken parallel to the transect $aa'$ (red line). They are
separated by a distance of 1.4~m. The elevation profiles of the same transects at different
times are used to study the dune instability.
}
\label{fig: ref}
\end{figure}

From April 2014 to November 2017, we performed a series of topographic surveys of the dunes
developing from the flat sand bed using a ground-based laser scanner Leica Scanstation C10
(Fig.~\ref{fig: ref}a). To compare these different measurements, a reference system of concrete
posts was installed over the entire experimental dune field (Figs.~\ref{fig: ref}b-d). For each
survey, the zone under investigation is scanned from four different view points. Over the 42
months of the experiment, the density of points varies from 472 to 2368 points/m$^2$.

To map surface elevation, we select only elevation points within the 4-sided polygon determined
by reference points t6, t7, t8 and t9  (Figs.~\ref{fig: ref}c-d). Within this polygon, to avoid
disturbances from the surrounding bedforms, we have chosen a central rectangular area with a width
of 48~m and a length of 82~m (ABCD in Figs.~\ref{fig: ref}c-d). The long side of this rectangle is
oriented Northwest-Southeast to align perpendicular to the final orientation of the dunes at the
end of the experiment. We remove the mean slope of this rectangular area by adjusting a plane to the
elevation data. This plane has always a southwest-facing slope during the entire duration of the
experiment. The residual topography is shown for different times in Fig.~5A of the main manuscript.
Within the rectangular area, we study 2D transects parallel to the main axis using a spacing of
1.4~m between two transects (see for example transect $aa'$ in Figs.~\ref{fig: ref}c-d). For each
of the 34 transects, we select the elevation data points in a 0.2~m wide band on either side. All
these points are horizontally projected on their respective transect line. After substracting the
average slope and the mean elevation, we resample these data to a regular spacing of 0.1, 0.25 and
0.35~m. Thus, we can use the fast Fourier transform method to explore different values within
the frequency domain.

Fig.~5B of the main manuscript shows the elevation profiles with respect to time for the transect $aa'$
shown in Figs.~\ref{fig: ref}c-d.  Over the 42 months of the experiment, the amplitude of the dunes
varies from a few centimeters to a few meters.

%
%
%
\subsection{Spectral analysis, amplitude and phase}
\begin{figure}
\centerline{
\includegraphics[width=1.00\linewidth]{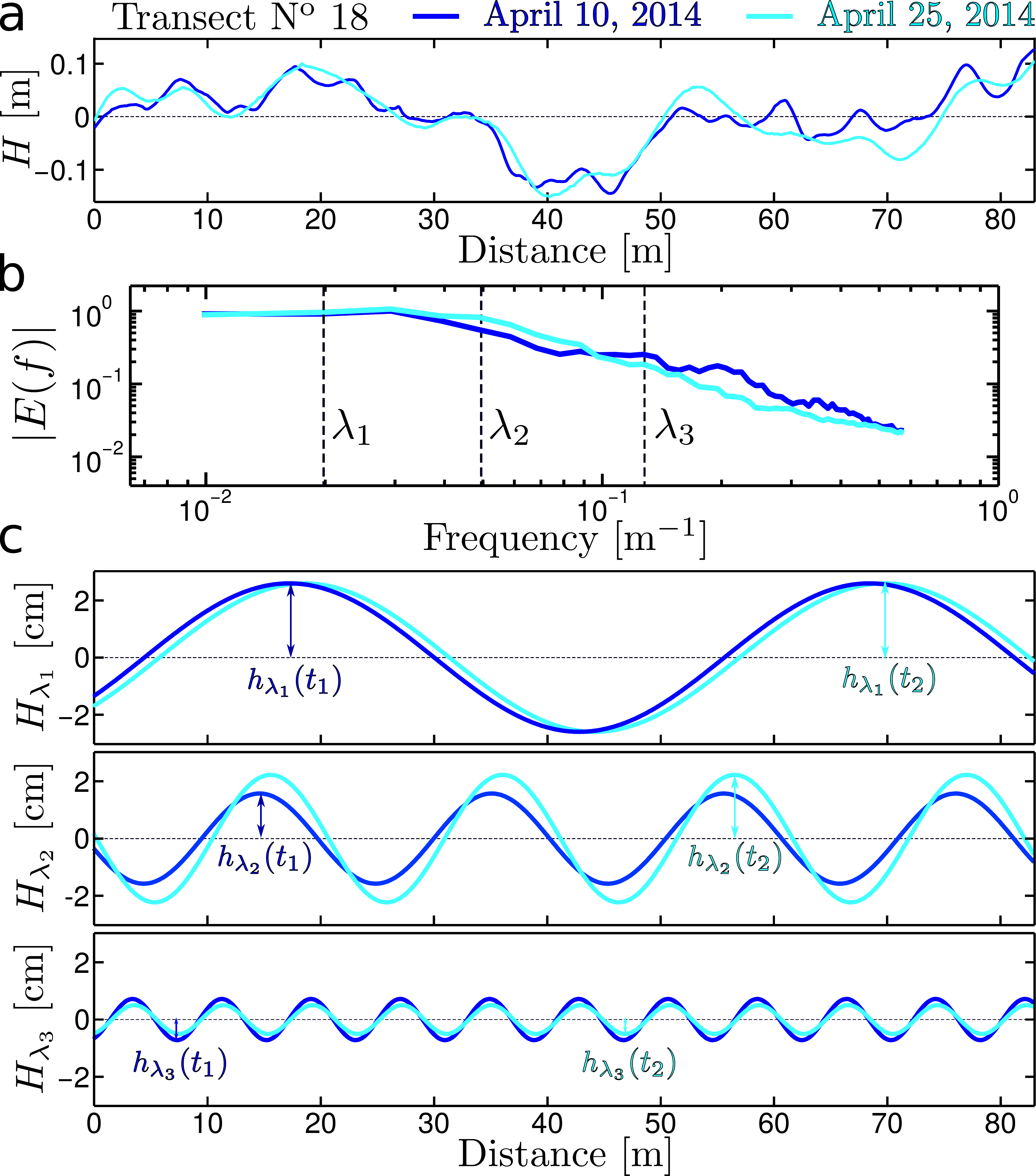}
}
\caption{
{\bf Spectral analysis of elevation profiles with respect to time}.
{\bf (a)}
Two consecutive elevation profiles of transect N$^{\rm o}$~18 during dune growth.
{\bf (b)}
Power spectral density of the elevation profiles using a fast Fourier transform. Dashed lines
show wavelengths $\lambda_{\{1,2,3\}}$ of $\{51.2 ,\, 20.5 ,\, 7.9 \}$ m.
{\bf (c)}
The reconstructed elevation profiles using individual wavelengths $\lambda_{\{1,2,3\}}$ and the
inverse Fourier transform. By definition, they take the form of sine waves
$H_{\lambda_i}=h_{\lambda_i}\cos(k_i x+\phi_{\lambda_i})$
where $k_i=2\pi/\lambda_i$ is the wave number and $\phi_{\lambda_i}$ the phase. Values of
$h_{\lambda_i}$ and $\phi_{\lambda_i}$ at different times can be used to estimate growth rate
$\sigma_{\lambda_i}$ and phase velocity $c_{\lambda_i}$, respectively.
}
\label{fig: spectral}
\end{figure}

For each transect and all topographic surveys, we use a fast Fourier transform method to analyze the signal
in the frequency domain (Figs.~\ref{fig: spectral}a,b). By selecting only one frequency value, the
individual contribution of each wavelength is computed by an inverse Fourier transform (Fig.~\ref{fig: spectral}c).
Thus, we obtain a sine wave
\begin{equation}
H_{\lambda_i}^j(t_n)=h_{\lambda_i}^j(t_n)\cos(k_i x+\phi^j_{\lambda_i}(t_n))
\label{eq: Hh}
\end{equation}
for each wavelength $\lambda_i=2\pi/k_i$ of each transect $j$ at the different times $t_n$ of the $n^{\rm th}$
topographic survey. Considering the 34 different transects, the amplitude $h$ and the phase $\phi$ can be used to
estimate growth rates $\sigma_{\lambda_i}$ and phase velocities $c_{\lambda_i}$ with respect to time, respectively.
We perform this analysis on the same elevation profiles sampled at different rates (0.1, 0.25 and 0.35~m) to verify
that it has no effect on the spectral behavior.

Taking as an example transect N$^{\rm o}$ 20 in the middle of the selected area, Fig.~\ref{fig: rates} shows
the logarithm of the amplitude
$A_{\lambda}^{20}$ of individual wavelengths with respect to the dimensionless time scale. A sudden change in
behavior is observed between $t_a=10.7$ and $t_a=11.3$, i.e. from October 30 to November 12, 2014. Over this time
period, there is an abrupt increase in growth rate and then a more irregular behavior than during the initial
phase. As shown in Fig.~5C of the main manuscript, it is also during this period that the mean amplitude of the
bedforms begins to increase more rapidly. These behaviors are interpreted as the transition between linear and
non-linear growth phases. This interpretation is supported by the observation of the first slip faces downwind
of the crest under the prevailing wind.

In order to quantify dune growth in the linear phase according to the dune instability (see Sec.~\ref{sec: tinst} and
Eq.~\ref{eq: exp}), we perform an exponential fit to the amplitude data from $t_a=0$ to $t_a=10.7$, i.e. from April 10
to October 30, 2014, for each wavelength and each transect (see red lines in Fig.~\ref{fig: rates}). The agreement
between the exponential regime and the data as well as the variation of the exponential rate at different wavelengths
indicate that there is a coherent behavior at different length scales, which can be analyzed thanks to a dispersion
diagram.

%
%
%
\subsection{Dispersion diagram of the growth rate}
\label{sec: sigk}
\begin{figure}
\centerline{
\includegraphics[width=0.95\linewidth]{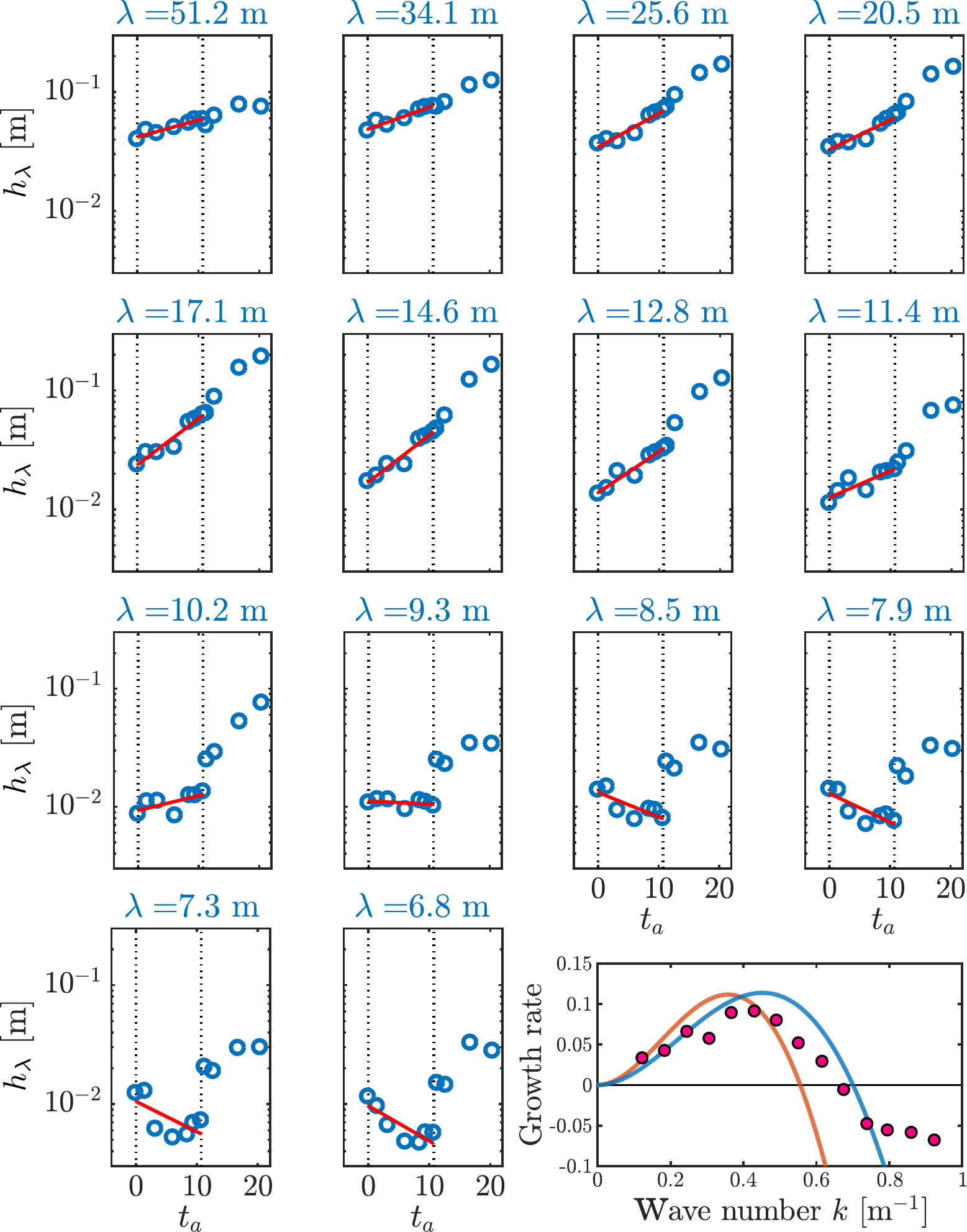}
}
\caption{
{\bf Growth rate of individual wavelengths}. Considering the elevation profiles of transect N$^{\rm o}$20
and a sampling rate of 25~cm, figures show the logarithm of the amplitude $h_{\lambda}$ of individual
wavelengths with respect to the dimensionless time scale; $t_a=0$ is set for April 10, 2014. Two
different regimes are observed before and after $t_a=10.7$, November 30, 2014. There are associated
with the linear and non-linear growth phases. The growth rate of each wavelength is determined by
an exponential fit performed during the linear phase ($0\leq t_a \leq 10.7$, dashed lines). As a
synthesis, the bottom right figure shows the dispersion relation, i.e. the growth rate with respect
to the wave number $k=2\pi/\lambda$. Solid lines are the same dispersion relations as in Fig.~6B of the
main manuscript: $\{l_{\rm sat},\, A,\, B\}=\{0.95,\,3,\,1.5\}$ (orange) and $\{0.7,\,1.96,\,0.96\}$
(blue). Note that small wavelengths have amplitude of less than one centimeter. These amplitudes are
comparable to the noise level of our topographic measurements, which prevents the  estimation of the
exponential decrease in growth rate for short wavelengths ($\leq$~8~m).
}
\label{fig: rates}
\end{figure}

Fig.~\ref{fig: rates} shows the dispersion relation for transect N$^{\rm o}$20 during the linear phase from $t_a=0$ to
$t_a=10.7$, i,e. from April 10 to October 30, 2014. In this figure, the exponential rates of growth or decay of the
different wavelengths $\lambda_i$ are plotted with respect to the wave number $k_i=2\pi/\lambda_i$. From the longest
wavelengths (i.e. smallest wave numbers), there is an  increase in the growth rate. The maximum growth rate is
reached for $k_{\rm max}\approx 0.43 ~{\rm m}^{-1}$ ($\lambda_{\rm max}\approx 14.6~{\rm m}$). For shorter wavelengths,
the growth rate is decreasing and the neutral mode is reached for a value of $k_0$ of about $0.67 ~{\rm m}^{-1}$
($\lambda_{\rm c}\approx 9.3~{\rm m}$). Then, shorter wavelengths have negative growth rates. These decay rates are
associated with small amplitudes of less than one centimeter, so that their rapid variations can not be captured
given the resolution of our topographic data.

\begin{figure}
\centerline{
\includegraphics[width=1.00\linewidth]{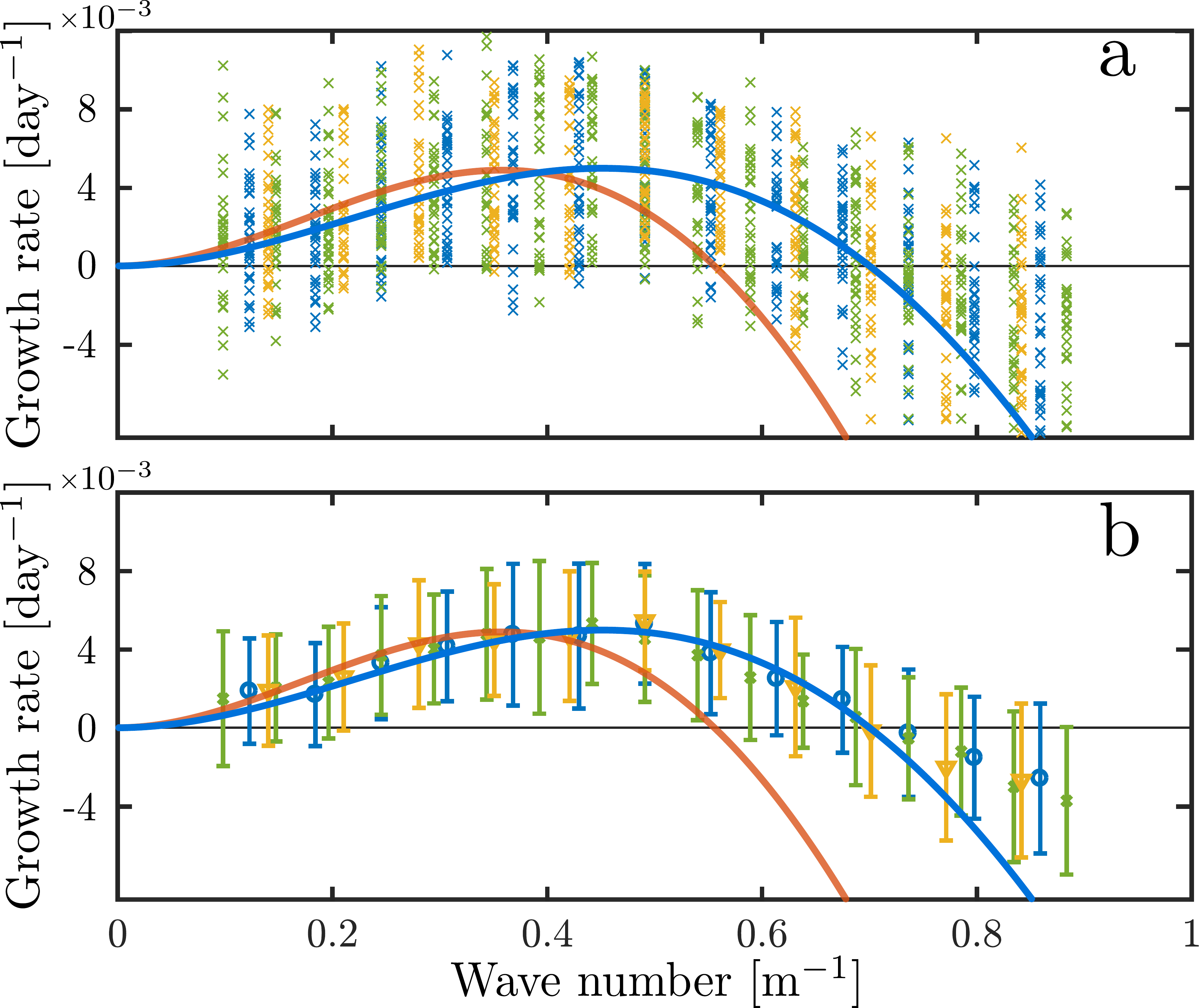}
}
\caption{
{\bf Dispersion diagram of the growth rate for all transects at different sampling rates}.
{\bf (a)}
Growth rates during the linear phase, i.e. from April 10, 2014 to November 30, 2014, with
respect to the wave number $k=2\pi/\lambda$ for each transect and sampling rates of  0.1~m (blue),
0.25~m (yellow) and 0.35~m (green).
{\bf (b)}
Mean and standard deviation of the growth rates averaged on all transects. Solid lines are the
same dispersion relations as in Fig.~6B of the main manuscript and in Fig.~\ref{fig: rates}:
$\{l_{\rm sat},\, A,\, B\}=\{0.95,\,3,\,1.5\}$ (orange) and $\{0.7,\,1.96,\,0.96\}$ (blue).
}
\label{fig: all_pts}
\end{figure}

Fig.~\ref{fig: all_pts}a shows the dispersion relations of the growth rate for all transects and the three sampling
rates. The mean and standard deviation of these growth rates are plotted with respect to the wave number in
Fig.~\ref{fig: all_pts}b. Using these data, we find
$k_{\rm max}\approx 0.45 ~{\rm m}^{-1}$ ($\lambda_{\rm max}\approx 14~{\rm m}$)
and
$k_{\rm c}\approx 0.7 ~{\rm m}^{-1}$ ($\lambda_{\rm c}\approx 9~{\rm m}$).

Theoretically, the growth rate writes
\begin{equation}
\sigma(k)=Q\,k^2\,\dfrac{B-A\,k\,l_{\rm sat}}{1+\left( k\,l_{\rm sat} \right)^2}
\label{eq: disp1}
\end{equation}
where $A$ and $B$ are the aerodynamic parameters (Sec.~\ref{sec: shift}), $l_{\rm sat}$ the saturation length
(Sec.~\ref{sec: lsat}), $Q$ a constant flux proportional to $u_*^2$. This equation does not take into account the
transport threshold and is therefore theoretically valid only within the limit of strong winds when the sand flux
is large. However, the role of the transport threshold is included in our estimates thanks to the dimensionless
time scale and a  prefactor correcting for the intensity of the sand flux. In practice, the growth rate derived
from the elevation profiles scales as $\sigma(k)/Q$ and we consider a prefactor
$\langle u_* \rangle^2/(\langle u_* \rangle^2-u_{\rm th}^2)$ that integrates all the variations of wind strength.
Thus, the parameter $A$ and $B$ in Eq.~\ref{taubprofile} and in Eq.~\ref{eq: disp1} are compatible with one
another.

The dispersion relation using Eq.~\ref{eq: disp1} and the values of $\{l_{\rm sat},\, A,\, B\}=\{0.95,\,3,\,1.5\}$
measured independently in the field are plotted in orange in Figs~\ref{fig: rates} and \ref{fig: all_pts} as in
Fig.~6B of the main manuscript. The blue curve in these figures corresponds to a dispersion relation with the
best-fit values of $\{l_{\rm sat},\, A,\, B\}=\{0.7,\,1.96,\,0.96\}$. In addition, to investigate the sensitivity
of the dispersion relation to the values of $\{l_{\rm sat},\, A,\, B\}$, the shaded area in Fig.~3c of the main
manuscript shows the best-fit values of $A$ and $B$ using values of $l_{\rm sat}\in [0.5;\,1.1]$ in
Eq.~\ref{eq: disp1}.  All these results underline the consistency of the experimental dispersion diagram derived
from the topographic data not only with the theory but also with our independent measurements of flow and
transport properties in the field.

%

%

%
\subsection{Dispersion diagram of the phase velocity}
During the linear-growth phase, the analysis of the variation of the phase $\phi(\lambda_i)$
(Eq.~\ref{eq: Hh}) as a function of time do not at present provide conclusive evidence about the
dispersion diagram of the phase velocity.
An obvious limitation comes from the time delay between two consecutive topographic surveys. Indeed,
the phase shift can be too large to be accurately computed, even by folding the data modulo the period from
large to short wavelengths. The orientation and magnitude of the resultant sand flux associated with
each time interval is another obvious issue. In fact, wind reversals and the subsequent back and
forth of the incipient bedforms were frequent during the experiment. Variation of the
phase data during periods of strong winds and small resultant sand flux are particularly difficult to
be interpreted.

\begin{figure}
\centerline{
\includegraphics[width=1.00\linewidth]{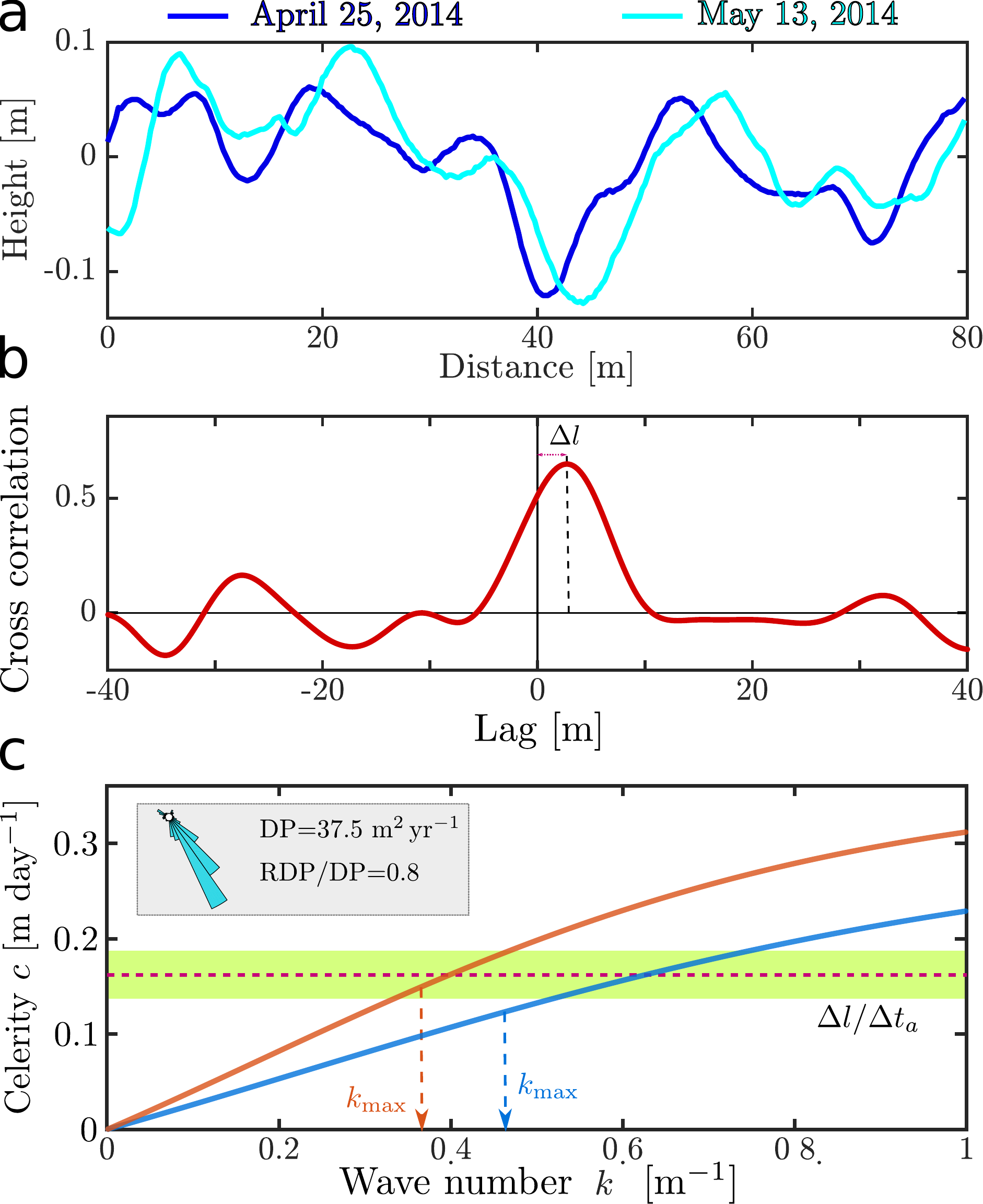}
}
\caption{
{\bf Dispersion diagram of the phase velocity}.
{\bf (a)}
Two consecutive elevation profiles of transect N$^{\rm o}$~16 during dune growth.
{\bf (b)}
The maximum value of the cross correlation diagram between the two elevation profiles
determines the dune migration distance $\Delta l=3.25~{\rm m}$ from April 25 to May 13, 2014
(i.e. $\Delta t_a=1.77$).
{\bf (c)}
Dispersion diagram of the phase velocity using the same parameters as in Fig.~6B of the
main manuscript and in Figs.~\ref{fig: rates} and \ref{fig: all_pts}b:
$\{l_{\rm sat},\, A,\, B\}=\{0.95,\,3,\,1.5\}$ (orange) and $\{0.7,\,1.96,\,0.96\}$ (blue).
Vertical dashed lines show the most unstable wavelength $\lambda_{\rm max}$ in both cases.
Using cross correlation, the horizontal dashed lines shows the mean migration velocity
$\Delta l/\Delta t_a$ averaged over all transects. The shaded area has a width of two
standard deviation using the same data. There is a quantitative agreement between
$c(\lambda_{\rm max})$ and the migration rate derived from $\Delta l/\Delta t_a$. This is
possible because during this period of time of the linear growth phase the wind regime is
almost unidirectional (see inset).
}
\label{fig: c_s}
\end{figure}

To simplify the problem, we can also estimate dune migration distance by cross correlation between two
elevation profiles over a time interval over which the wind is almost unidirectional, from April 25
to May 13, 2014, (Figs.~\ref{fig: c_s}a,b). The dune migration rate can then be computed using the
dimensionless time scale to be compared to theoretical dispersion relations. With the same notation
as in Eq.~\ref{eq: disp1}, dispersion relations write
\begin{equation}
c(k)=Q\,k\,\dfrac{A+B\,k\,l_{\rm sat}}{1+\left( k\,l_{\rm sat} \right)^2}.
\label{eq: disp2}
\end{equation}
Considering the dimensionless time scale and the same prefactor as for the growth rate, Fig.~\ref{fig: c_s}c
shows the dispersion relations of the phase velocity using the same parameters as in Figs.~\ref{fig: rates}
and \ref{fig: all_pts}. It also shows the phase velocity of the most unstable wavelength $\lambda_{\rm max}$,
the mean value of the migration rate averaged over all transects and the corresponding standard deviation.
As expected, the migration rate derived from the cross correlation is close to the phase velocity of the
most unstable wavelength.

%


\bibliographystyle{naturemag}
\bibliography{biblio}